\documentclass[12pt,reqno]{article}
\usepackage[utf8]{inputenc}
\usepackage[round]{natbib}
\usepackage{color}
\usepackage{array}
\usepackage{float}
\usepackage{url}
\usepackage{amsmath}
\usepackage{amsthm}
\usepackage{amssymb}
\usepackage{graphicx}
\usepackage{booktabs}
\usepackage{enumitem}
\usepackage{setspace}
\usepackage{tabularx, booktabs}
\usepackage{threeparttable}
\usepackage{appendix}
\newtheorem{hyp}{Hypothesis}

\newtheorem{innercustomres}{Result}
\theoremstyle{remark}

\theoremstyle{definition}
 
 \theoremstyle{definition}
 
\theoremstyle{definition}

\theoremstyle{plain}

\theoremstyle{plain}

\theoremstyle{plain}

\theoremstyle{plain}

\theoremstyle{remark}
\newtheorem*{claim*}{Claim}
\newcommand{\sym}[1]{{#1}}
\usepackage{geometry}
\usepackage{hyperref}
\hypersetup{
    colorlinks=true,
    linkcolor=red,
    citecolor=blue,
    filecolor=magenta,      
    urlcolor=cyan
    }

\begin{document}

\title{Strategic Responses to Personalized Pricing and Demand for Privacy: An Experiment\thanks{We thank Svenja Hippel, Renke Schmacker, Johan Stennek, and the seminar and conference audiences at the 15th Nordic Conference on Behavioral and Experimental Economics, University of Bonn, and the Lisbon Game Theory Meetings for comments and suggestions. Inácio Bó acknowledges financial support from National Natural Science Foundation of China key project (no. 72033006). Li Chen acknowledges financial support from Tore Browaldhs stiftelse BFh18-0007.   Rustamdjan Hakimov acknowledges financial support from the Swiss National Science Foundation (Project number 100018\_207722)}}

\author{Inácio Bó\thanks{University of Macau.
Avenida da Universidade, Taipa, Macau, China. Email: inaciobo@um.edu.mo}, \,
Li Chen\thanks{Corresponding author. Tongji University, Siping Road 1239, Shanghai 200092, China; University of Gothenburg, Vasagatan 1, Box 640, Sweden.  Email: chen\_li@tongji.edu.cn}, \,
Rustamdjan Hakimov\thanks{University of Lausanne, Internef 536, Quartier de Chamberonne, CH-1015, Lausanne, Switzerland. Email: rustamdjan.hakimov@unil.ch}
}

\date{November, 2024}
\maketitle
\begin{abstract}
We consider situations in which consumers are aware that a statistical model determines the price of a product based on their observed behavior. Using a novel experiment varying the context similarity between participant data and a product, we find that participants manipulate their responses to a survey regarding personal characteristics, and manipulation is more successful when the contexts are similar. Moreover, participants demand less privacy, and make less optimal privacy choices when the contexts are less similar. Our findings highlight the importance of data privacy policies in the age of big data, in which behavior in apparently unrelated contexts might affect prices.
\end{abstract}

\noindent \textit{JEL. classification}: L11, C91, D91, M38

% % \end{frontmatter}
\clearpage

\section{Introduction\label{section:Introduction}}

Advances in information technology in recent decades have led to an explosive growth of consumer data. Firms can exploit these data for more accurate target decisions like pricing. 
Despite increasing consumers' awareness of privacy protection and government regulations such as the General Data Protection Regulation (GDPR) in the European Union (EU), consumers' ability to make informed choices regarding their privacy is often compromised due to incomplete information regarding what data is collected and how that data is used, particularly in complex ``big data'' environments. 

This paper studies one such complex environment with personalized pricing. Personalized pricing uses information on each individual's observed characteristics to implement consumer-specific price discrimination. Tracking tools, such as cookies, enable firms to build profiles of consumers on the Internet and target them with individualized prices, thus extracting consumer surplus.
For example, airline companies and car rental services are known to sell products and services online according to user features, such as location.\footnote{
Finding systematic evidence on personalized pricing is challenging, consumer reports such as \cite{schleusener2016expertise} and \cite{economics2018consumer} shed some light on personalized pricing practices in Germany and EU countries.
These reports reveal that in addition to location, airline companies charge higher prices for people using Mac operating systems and mobile phone devices. Personalized pricing practices have also received media attention. See, for example, a BBC report on rental services,  \url{https://www.bbc.com/news/business-28756674}, last accessed on August 10, 2022. Recently, the Dutch Authority for Consumers and Markets found that an online shopping platform, Wish, used personalized pricing based on locations in the EU,  \url{https://www.acm.nl/en/publications/following-acm-actions-wish-bans-fake-discounts-and-blocks-personalized-pricing}, last accessed on April 12, 2023. 
} 
With growing awareness of personalized pricing, consumers can, occasionally, take countermeasures.\footnote{A recent European Commission consumer study has reported a growing awareness of personalized pricing among consumers in EU countries \citep{economics2018consumer}. According to the survey, 44\% of consumers know about personalized pricing and claim they understand how it works. A similar share of respondents believes that personalized pricing offers them discounts and reductions and provides them with the best available price.} In the case of airline tickets or car rental services, they can protect their privacy by deleting cookies or using the ``private browsing'' option. Alternatively, they can respond strategically to cut a nice bargain by, for example, changing their locations through a virtual private network (VPN), provided that they know how prices are set according to locations.\footnote{Multiple websites and forums discuss the best strategies to avoid higher prices for airline tickets using such techniques. For example, \url{https://www.makeuseof.com/tag/insanely-cheap-flights-vpn/}, last accessed on August 10, 2022.} 
While it might be easy to guess the ``cheap'' locations, the link between consumer information and firms' pricing models is becoming increasingly blurred with the use of big data. Big data allows firms to better explore the link between the willingness to pay (WTP) for a product and consumer characteristics beyond these easily observed ones.\footnote{Web tracking services often provide not only basic information regarding visitors such as locations, age, and gender, but also additional information, such as their interests and tastes in movies, music, and sports. The marketing literature has recognized the value of using personality traits, values, lifestyles, and emotions---known as psychographics---for consumer targeting \citep{gunter2014consumer}.} This data might be less direct, but it can capture additional consumer information correlated with consumers' WTP and is cheap to obtain \citep{organisation2015data}.  
It is essential to understand how such technologies affect consumer behavior and welfare. In particular, we would like to know, when the link between consumer data and the pricing model is less obvious, how consumers weigh between privacy protection and ``gaming the system'', and  whether they can respond strategically in their own favor. However, identifying these choices in field data is difficult. The ideal data would need to determine whether a specific consumer is aware of personalized pricing and disentangle strategic from non-strategic responses.  
 
In this study, we experimentally analyze consumers' responses---how they report and manage their data---to personalized pricing. We identify whether they are aware of personalized prices and discern their strategic responses in a controlled online experimental setup, using a sample from the US population through the Prolific platform. The product for sale is a lottery with a 50\% probability of winning £5.\footnote{Since Prolific is a UK-based company, and at the time of the experiment this was the only currency used for payments, all rewards were fixed in British pound sterling. Subjects' payments were converted to U.S. dollars using the payment day's exchange rate.} A key element in our design is the link between participants' survey responses and the predicted WTP that determines the personalized prices. To vary the degree of similarity between the survey context and the product, we employed two surveys to predict WTP. The first survey consists of questions commonly used by insurance firms or banks for risk profiling (Risk treatment), thereby making the link between the responses and the predicted WTP for the lottery relatively straightforward, as both share the same context---risk. The second survey asks participants to rate various movie genres (Movies treatment), thereby making the relationship less obvious. Nevertheless, movie ratings could connect to the WTP for the lottery through channels such as personality and gender, which in turn correlate with risk preferences \citep{rentfrow2011listening,becker2012relationship,croson2009gender}. For example, the preference for horror movies is correlated with thrill-seeking behavior, which in turn correlates with risk-loving preferences.

The experiment consists of two stages.
In the first stage, we collected responses to these two surveys and the WTP for the lottery through multiple price lists (MPL).\footnote{In an MPL task, participants decide whether they would purchase a lottery ticket at various price points, ranging from very low to very high. The prices at which participants switch from buying to not buying the ticket enable the estimation of their WTP.} This information was used to train a pricing model.  There is no strategic aspect at this stage: participants received a fixed payment for answering the surveys. In the main experiment stage, participants went through one of the surveys, depending on their treatment group, and were informed that the survey responses would be used to determine the price by a statistical model trained with real data. There is a strategic aspect at this stage: their responses could affect their payments. However, participants were allowed to select a \textit{privacy option}, in which they paid a cost to hide their survey responses after submitting them, but before observing the suggested price. In this case, an anonymous price was offered instead.

 \textit{First}, we show that participants manipulate their responses in both treatments, but more so in Risk than in Movies. Comparing these responses with the training data, we observe a significant difference between the treatment and training data in seven out of ten Risk survey questions and two out of ten Movies survey questions.\footnote{We consider three dimensions---mean, distribution, and variance---and interpret a significant difference in at least one of these dimensions as a sign of manipulation in a question.} \textit{Second}, manipulation by participants is more successful in Risk. In other words, the predicted WTP in Risk is significantly lower than that in the training data. We find no significant differences in the predicted WTP between the training data and Movies. Consequently, individualized prices are significantly lower in Risk than in Movies. \textit{Third}, contrary to our prediction, participants are significantly more likely to buy the privacy option in the Risk treatment than in the Movies treatment. The decisions to buy the privacy option are significantly more often optimal in Risk than in Movies. The difference is mainly driven by those who do not buy the privacy option when they should, as the individualized price is higher than the anonymous price. 
These behavior patterns are consistent with participants being na\"{i}ve regarding the relevance of their responses to movie ratings for personalized pricing. 
As a result, participants have significantly higher payoffs in Risk than in Movies.

Our main contributions are twofold. First, we demonstrate that consumers can be strategic against pricing algorithms, but the success of strategic manipulation is substantially reduced when the pricing model relies on data from less related contexts. Although this may not be surprising, there has been a lack of causal empirical evidence that supports both strategic responses to pricing algorithms and consumers' diminished performance in more complex setups, which are likely to become prevalent in the era of big data.

Second, while providing privacy options is essential, somewhat counter-intuitively, privacy is less demanded when pricing models are complex. This result suggests that consumers may be naïve regarding the usefulness of data from less related contexts for personalized pricing, which could potentially decrease consumer welfare due to incorrect strategic responses and less optimal sorting into private browsing. 
As big data becomes more prevalent, our results call for interventions to increase awareness regarding the scope of pricing models and promote policies that enhance transparency regarding the inputs used in algorithms and their significance for price discrimination.

\textbf{Related literature.} Our findings are related to the literature on the welfare implications of targeted price discrimination.
The extant theoretical literature primarily considers behavior-based price discrimination---where firms offer different prices conditional on the histories of consumer purchases---and analyzes the optimal pricing strategies by firms, assuming consumers are not strategic \citep{villas1999dynamic,fudenberg2000customer,choe2018pricing}. The results are surveyed in \citet{fudenberg2006behavior} and \citet{acquisti2016economics}. A recent study by \cite{rhodes2024personalization} considers broader personalized pricing that, similar to our setup, uses information on consumer characteristics. Their theoretical analysis focuses on the externality of privacy choices, which can reduce overall consumer welfare. 
A smaller set of papers have considered the possibility of consumers' strategic effort to influence the price, mostly assuming consumers can successfully strategize through their purchasing behavior \citep{taylor2004consumer,acquisti2005conditioning,chen2017competitive}. One exception is \cite{bonatti2020consumer} who show, in a theoretical model, that information is of significance to consumer behavior under price discrimination. Specifically, they examine the welfare of using aggregated consumer information, such as credit scores, for price discrimination and find that price discrimination based on purchase histories harms na\"{i}ve consumers who ignore the link between current purchases and future prices, while benefiting sophisticated consumers who understand how firms use their scores for pricing and, crucially, when they can observe their scores. Sophisticated consumers who do not observe their scores can still be harmed. Our experimental results that indicate worse consumer welfare in Movies than in Risk provide an empirical contribution to their theoretical result of lower welfare for sophisticated consumers when they do not observe their scores. Like theirs, we also study a monopolist setup, but consider price discrimination based on consumer characteristics rather than purchase histories, which is easier to implement in experiments. 

Despite the large theoretical literature, empirical evidence on targeted price discrimination is limited. Moreover, evidence on strategic behaviors is even more scarce, as it is difficult to identify them in the field. 
A few papers use field data to estimate the impact on consumer welfare---without considering the strategic responses---of offering prices based on past histories \citep{shiller2020approximating} and observable characteristics \citep{waldfogel2015first,dube2017competitive,dube2019personalized}. They find that price discrimination generally increases consumer welfare, although the magnitude depends on the method used. In our experiments, price discrimination is designed to yield higher revenue than in the scenario without it. This enables us to analyze differences between contexts with different levels of similarity. Our study complements the existing empirical literature by providing causal evidence that the similarity of context affects consumers' strategic decisions on privacy. 
  
The strategic responses explored in our paper connect to the literature on consumer attitudes and behaviors in various contexts. Closely related to our paper, \citet{hagenbach24} conduct an experiment where participants first answer six questions about themselves. Then, participants can choose to hide some of their answers to prevent the algorithm from guessing their response to a target question. The algorithm relies on correlations between the answers. The results show that participants fail to optimally conceal their answers from the algorithm, allowing for profiling. Additionally, the study finds that providing explanations of how the algorithm works actually decreases the rate of optimal disclosure, underscoring the complexity of strategic responses.
\cite{leibbrandt2020behavioral} uses experiments to study when firms would use price discrimination when consumers can be averse to such action. The results align with a reference point model in which consumers care about fairness and redistribution. In a field experiment, \citet{dargnies24aihr} show that job candidates attempt to strategically manipulate their responses to a survey on behavioral traits in the context of algorithmic hiring. 
When it comes to privacy, research has revealed that consumers' decision-making is affected by behavior biases---such as immediate gratification \citep{acquisti2004privacy} and status quo bias \citep{john2011strangers}---and that context matters for the WTP for privacy \citep{tsai2011effect,beresford2012unwillingness,jentzsch2012study}. In another related paper, \cite{hillenbrand2019strategic} explore consumers' strategic behavior in a price discrimination setting in which both product fit and price are crucial. The strategic incentive arises because consumers' search behavior could reveal information that leads to higher prices. Their experimental results uncover a different type of welfare loss associated with improved consumer data under price discrimination: consumers, 
fully aware that their information is used for price discrimination and could learn about pricing over time, may end up receiving a worse product fit in exchange for a potentially lower price. 

Our findings also relate to strategic mistakes observed in the market design and matching literature and add to the debate on the design of privacy regulation policies. In our setting, certain participants manipulate their responses even though they should not, thus leading to reduced welfare. This type of mistake is similar to those observed in the context of matching students to schools and colleges, where seemingly innocuous design details can prevent students from adopting optimal strategies when faced with complex environments  \citep{hassidim2017mechanism,shorrer2017obvious,chen2015self}. \citet{boldrini2024} show that when the features of the algorithm used by Amazon to recommend sellers are explained, consumers are less likely to use it, thereby suggesting that consumers are na\"{i}ve when using algorithmic recommendations on platforms, which is similar to what we observe in our context. 

Current regulations on privacy, such as GDPR, rely on ``notice and consent'' for data collection, without much emphasis on how the data is actually used. Our results suggest that the sole use of notice and consent is probably insufficient to protect consumers due to strategic mistakes. Additional awareness should be raised and guidance be provided on the nature of the relationship between their data and how it is used so that consumers can manage their privacy.

The remainder of this paper is organized in the following manner: 
We present our experimental design in Section \ref{section:design} and outline our hypotheses in Section \ref{section:hypotheses}. We present our findings in Section \ref{section:results}, and conclude and discuss potential policy implications in Section \ref{section:conclusion}. We provide omitted information in the Appendix.

\section{Experimental Design}\label{section:design}

The primary goal of our experiment is to identify subjects' strategic responses to personalized pricing and their demand for privacy.
This is achieved through a two-stage experiment that involves survey questions that are both more related and less related to the product for sale---a lottery with a 50\% probability of winning £5, and decisions to purchase the lottery at a given price. We selected lottery as a product, as we could leverage well-established survey questions designed to measure risk preferences for our product-related survey questions. This approach enables us to establish a direct link among subjects' responses, their risk preferences, and WTP for the lottery, thereby facilitating a clear interpretation of our results. Moreover, previous research indicates a significant degree of heterogeneity in risk preferences among individuals \citep{rieger2015risk,vieider2015common}, which provides a good variation for our experiment. 

A key difference between the two stages is how subjects were incentivized, which we will describe shortly. At the beginning of the experiment, we asked subjects for their gender, age, and consent to participate in the study. Gender and age are subsequently used as control variables throughout our analysis. 
We deployed our experiments on the Prolific Platform with a sample from the US population.  
The design and main hypotheses are registered in the AEA RCT registry (AEARCRT-0009440).\footnote{The pre-registration includes two additional treatments---ScopeRisk and ScopeMovies. These treatments inform subjects regarding the range between minimum and maximum personalized prices. Our interest in the effect relied heavily on the assumption that subjects underestimate the scope of price discrimination, which is not the case in our data. We opted out of presenting the data from these treatments in the main text to simplify the paper's exposition and motivation, thereby focusing on the difference between the Risk and Movies surveys. We present the results of these treatments in the Online Appendix.} The design received ethical approval from the Ethics Board of HEC Lausanne. Details of the experiment instructions and survey questions are provided in the Appendix \ref{section:Appendix_instructions}.

\subsection{Training Sample}
 In the first stage, we collected data, which was then used as a training sample for developing a personalized pricing model (or algorithm). Subjects in this stage received a fixed payment for their responses and, therefore, their participation did not affect the price. Consequently, we interpreted their responses as truthful.  
 We collected responses to the following two surveys in a random order for our training sample:

\begin{enumerate}
  \item A survey to identify the risk preferences of subjects, similar to the assessments conveyed by insurance companies (Risk survey). 
  \item A survey where subjects rate movie genres (Movies survey). 
\end{enumerate}

Table \ref{tab:riskQuestions} outlines the questions in the Risk survey, along with the available responses and their associated values. Responses were constructed so that those with larger values indicate stronger preferences for risk-taking. In the Movies survey, subjects provided ratings for various movie genres---such as Romance, Horror, and Action---on a scale from 1 to 10. A rating of 1 indicates the lowest preference, while a rating of 10 signifies the highest preference. 

\begin{table}[htp]
\centering
\caption{Questions in the Risk survey}\label{tab:riskQuestions}
\begin{threeparttable}
\footnotesize
% Increase the vertical padding
\renewcommand{\arraystretch}{1.5}
% Increase the horizontal padding
\setlength{\tabcolsep}{10pt}
\noindent
\begin{tabularx}{\textwidth}{rX|X}

\toprule
\multicolumn{2}{c}{\textbf{Question}} & \multicolumn{1}{c}{\textbf{Responses and associated values}}\\
\hline
R1 & I am prepared to forego potentially large gains if it means that the value of my investment is secure & Strongly Agree (1) to Strongly Disagree (5) \\
\hline
R2 & Over the next several years, you expect your annual income to & Decrease substantially (1) to Grow substantially (5) \\
\hline
R3 & Imagine that due to a general market correction, one of your investments loses 14\% of its value. What do you do? & Sell (1), Hold (2), Buy more (3) \\
\hline
R4 & What is the current amount of insurance you buy (life insurance, home insurance, etc)? & Much more than most people I know (1) to Much less than most people I know (5) \\
\hline
R5 & Assuming you are investing in a stock, which one would you choose? & Stable corp. with dividends (1) to High potential start-ups (3) \\
\hline
R6 & Have you ever borrowed money for the purpose of making an investment? & No (1), Yes (2) \\
\hline
R7 & You have just reached the \$10,000 plateau on a TV game show. Which do you choose? & Take \$10,000 (1) to 5\% chance for \$100,000 (4) \\
\hline
R8 & Do you smoke cigarettes? & No (1) to Yes, daily (3) \\
\hline
R9 & In an amusement park, which describes your type best? & No adrenaline (1) to Extreme attractions (3) \\
\hline
R10 & Which describes your preferences for future employment best? & Stable government job (1) to Self-employed (3) \\
\hline
\end{tabularx}
\begin{tablenotes}
    \footnotesize
\item Notes: This table presents a shortened version of the questions, responses, and associated values from the Risk survey. The full details are available in the Appendix \ref{section:Appendix_instructions}. 
\end{tablenotes}
\end{threeparttable}
\end{table}

After completing the surveys, subjects entered the final stage, where we elicited their WTP for the lottery using MPL. Each row of the lists presents subjects with the option to choose between buying the lottery or not buying it at a specific price. The price varied from £0.60 to £4 in £0.20 increments.\footnote{Our price range does not cover the entire spectrum due to our focus on capturing realistic WTP values relevant for model training. Referring to estimates of relative risk aversion from \cite{holt2002risk}, we acknowledge that our range does not allow for the precise estimation of WTP for the top 1\% of risk-loving participants and the 13\% of the most risk-averse participants.} One of the rows was selected randomly, and the associated choice of buying or not was implemented and paid out for 20 randomly selected subjects.\footnote{While our payment scheme incentivized approximately 2.5\% of participants in the training sample, we do not expect it to alter responses relative to full incentivization. \citet{holt2002risk} find that estimates of risk aversion in low-stakes scenarios are similar for both incentivized and non-incentivized decisions. \citet{charness2016experimental} found no substantial differences in responses when incentivizing only 10\% of subjects, and \citet{ahles2024testing} similarly reported no significant effects when comparing 10\% and 1\% probabilistic payments. Our approach aligns with common practices for surveys on Prolific, where probabilistic payments are regularly used \citep{hvidberg2023social,andre2023fighting}.}

\subsection{Treatments: Risk and Movies}

In the main experiment stage, we ran between-subjects treatments in which subjects completed one of the surveys. ``Risk'' refers to the treatment with the Risk survey. ``Movies'' refers to the treatment with the Movies survey.

Prior to the survey, subjects were \textit{informed} that a statistical model built using answers from real subjects would determine the lottery price in a subsequent round of the experiment. This was the only information that subjects were given about the pricing model. 
To ensure subjects could afford the lottery, which had a maximum price of £2.09 according to our pricing model, we paid £2.20 to subjects for completing the survey. 

After completing the survey but before observing the price, subjects were allowed to choose to conceal their survey responses from the seller (imitating a privacy---or private browsing---option) for a cost of £0.10. If they decided to hide the survey responses, the price would be the one that maximized revenue, given the distribution of the WTP in the entire training sample. We referred to it as the \textit{anonymous price}. To avoid curiosity motives, subjects were informed that they would learn about both the individual and the anonymous prices at the end of the experiment.

In the following step, subjects had to decide whether to buy the lottery at a given price $p$, which was either determined by the algorithm using their survey responses (if the subjects decided not to hide their responses) or by anonymous pricing. When they chose to buy, the lottery was played out, and the subject's payoff for the last round was £$5-p$ if won, or $-p$ if lost. In cases where the payoff was negative, it was deducted from the reward given for filling out the survey.

In the last experimental task, we elicited the subjects’ beliefs regarding the lowest and the highest individualized prices that the pricing model could generate. We paid them £0.10 if they were within £0.20 of the lowest price and £0.10 if they were within £0.20 of the highest price.

\subsection{Sample data}
\label{subsec:sampleData}

We collected 804 responses for the training data, using a sample representative of the U.S. population. The average duration of the survey in the training sample was 5.5 minutes. The participation payment was £1, and in addition, 20 randomly selected respondents received their payoff from the lottery task.
We collected 302 and 301 responses in the Risk and Movies treatments, respectively.\footnote{For these treatments, the sample was not representative of the U.S. population, but was gender-balanced. In section \ref{subsec:robustnessAndData}, we indicate the reasons why we don't believe that this affects our analysis.}  The average duration for these was 7.5 minutes. The average payoff of the participants was £6.30, including a payment of £0.75 for participation.

\section{Hypotheses}\label{section:hypotheses}

Our first hypothesis relates to whether subjects' behavior indicates that they make clear attempts to strategize their survey answers in response to the knowledge that their answers will determine the prices they face.

\begin{hyp}
Participants attempt to manipulate the responses: The distributions of responses in Risk and Movies are significantly different from the responses to the same questions in the training data.  
\end{hyp}

Next, we hypothesized that subjects have a better understanding of the relationship between demand for lotteries and a survey on risk preferences, as opposed to one on movie genre preferences. In principle, it is conceivable that a pricing model based on the Movies survey would result in a wider range of manipulation possibilities. If subjects know how to work with them, they could be more successful in their attempts in obtaining lower prices.
However, we conjecture that the subjects will fail to successfully infer the more complex relationship between their WTP and movie preferences.

\begin{hyp}
Participants are more successful in strategically shifting down their predicted WTP in the Risk survey as compared to that in the Movies survey : The predicted WTP in the training data is significantly higher than in the Risk survey, and is not significantly different from that in the Movies survey. 
\end{hyp}

While subjects may not be able to manipulate the Movies survey as well as the Risk survey, we conjecture that they will be aware of the complexity involved in the former and, consequently, choose the privacy option more often in the Movies treatment. This is because they may wish to avoid the risk of making incorrect choices.

\begin{hyp}
Participants anticipate the complexity of manipulating the Movies treatment and select the privacy option significantly more often than in the Risk treatment. 
\end{hyp}

Additionally, a clearer understanding of the relationship between their responses and the predicted values in the Risk treatment is expected to result in a higher proportion of optimal choices for the privacy option---that is, choosing the privacy option if the price under privacy is lower than the personalized price.

\begin{hyp}
The proportion of optimal privacy choices is higher in the Risk treatment than that in the Movies treatment.
\end{hyp}

\section{Results}\label{section:results}
\subsection{The pricing model}
\label{subsec:pricingModel}

\begin{figure}
  \centering
  \includegraphics[width=0.8\textwidth]{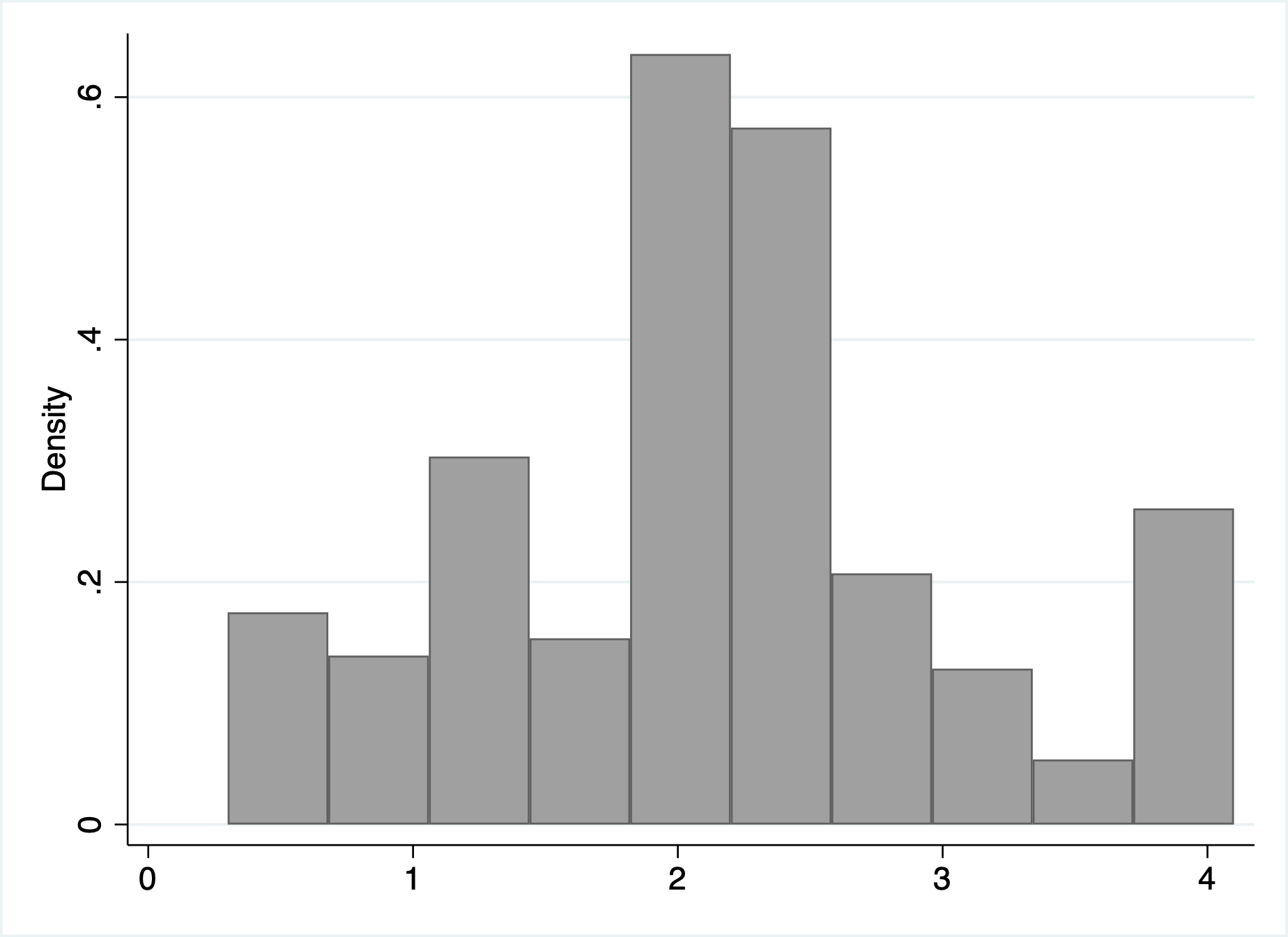}
  \caption{Distribution of the WTP in the training data}
  \label{fig:wtpdistr}
\end{figure}

This subsection presents the descriptive results of the training data and the development of the pricing model for our main treatments.
First, we define the WTP as the switching point of the choices between lotteries in the MPL. From among 804 collected responses, 67 had multiple switching points, and we excluded them from the analysis, as we would have had to make additional assumptions to assign a WTP to these subjects. For the remaining 737 subjects,\footnote{Note that some subjects did not answer one or several survey questions. Thus we cannot use them in the model estimation. This leaves us with a sample of 731 subjects in Risk and 723 in Movies.} we calculated the WTP value as the average between the prices of the lotteries adjacent to the switching point from not buying to buying. For example, if a subject chose not to buy a lottery for £2.0 but chose to buy it for £1.8, we assigned a WTP of £1.9 to this subject.  For those subjects who always chose not to buy the lottery, even at a price of £0.60, we assigned a WTP of £0.30. For the subjects who always chose to buy the lottery, even at a  price of £4, we assigned a WTP of £4.10. Figure \ref{fig:wtpdistr} presents the distribution of the resulting WTP. A simple optimization reveals that the revenue-maximizing price is £1.85. In the training sample, this price would result in 521 sales, with a total revenue of £963.85.

Prior to developing a predictive model, we present the correlations between the WTP and answers to each survey. Table \ref{tab:correl} presents the Spearman correlations of the WTP with the responses to the Risk survey (Panel A) and the Movies survey (Panel B). We observe a significant correlation of the WTP with seven out of ten Risk survey questions at the 5\% level. 

As Table \ref{tab:riskQuestions} shows, the values associated with the answers for the Risk survey are constructed such that larger values are associated with a stronger preference for risk-taking. Therefore, higher values for their responses should be correlated with higher WTP for a lottery. In our experiment, the WTP correlated the most with the question regarding forgoing gains for securities of investment and game show decisions, and the correlations are in the predicted direction. Specifically, the subjects with a higher WTP tended to disagree more with the statement ``I am prepared to forego potentially large gains if it means that the value of my investment is secure." In addition, subjects who chose to bet on alternatives with different winning probabilities and awards in a hypothetical TV game show scenario, rather than quitting with a safe option, tend to have a higher WTP for the lottery. As for the Movies survey, the correlations between the responses and WTP are more complex and only five out of ten questions significantly correlate with it. The strongest correlation is with the ratings for crime movies. Thus, our hypothesis that survey responses might correlate with the WTP found support in the data. 

As mentioned earlier, we based our assumption that responses to the Movies survey would correlate with WTP on two channels. First, the big five personality traits correlate with movie ratings \citep{rentfrow2011listening} and risk \citep{becker2012relationship}. Second, through gender, women are more risk-averse than men \citep{croson2009gender}. We can verify the latter channel by testing whether movie ratings predict the gender of the respondent. Indeed, in our training data, simple ordinary least squares (OLS) regression of gender on the movies ratings has an adjusted $R^2$ of 24\%, thereby resulting in the correct categorization of the subjects' gender in the linear discriminant analysis for approximately 74\% of subjects (see Tables \ref{tab:AppFemalemov} and \ref{tab:AppFemalemovDisc} in Appendix \ref{section:AppendixA}).

\begin{table}[htp]\centering
\caption{Spearman correlations of WTP and survey answers }\label{tab:correl}
\footnotesize
\begin{tabular}{lr|lr}\toprule
\multicolumn{2}{l}{Panel A}&\multicolumn{2}{l}{Panel B}\\
Risk survey &WTP& Movies survey &WTP\\\midrule
R1: Forgo gains for secure investment &0.12***&M1:Romance&0.04 \\
R2: Annual income&0.08**&M2:Horror&0.10**  \\
R3: Loss of 14\%, action &0.08** &M3:Action&0.09** \\
R4: Current insurance amount &0.07** &M4:Documentary&-0.01 \\
R5: Which stock you choose &0.00 &M5:Foreign&-0.01 \\
R6: Borrow for investment &0.09** &M6:Fantasy&0.06 \\
R7: Gameshow 10k safe vs alternative &0.14*** &M7:Comedy&0.11*** \\
R8: Smoking &0.05 &M8:Historical&0.01 \\
R9: Amusement park &0.05 &M9:Crime&0.16*** \\
R10: Future employments &-0.07** &M10:Thriller&0.09** \\
\bottomrule
Notes: * $p<0.10$, ** $p<0.05$, *** $p<0.01$.
\end{tabular}
\end{table}

The first step in developing the pricing model was to create a predictive model of WTP using survey answers. There are numerous ways to address this question, including the use of machine learning techniques. However, we chose a simpler model, as our primary objective was not to achieve the highest level of precision. Instead, we aimed to develop a straightforward model that can be easily implemented in Qualtrics and enables a clear interpretation of the coefficients. We used the OLS of the WTP for all variables of a survey and all possible pairwise interactions among these variables. We then eliminated all variables with $p$-values above 0.5, then 0.3, and then 0.1. The resulting model was used to predict the WTP. Tables \ref{tab:Appwtpstand} and \ref{tab:Appwtpmov} in Appendix \ref{section:AppendixA} present the resulting models. 
As expected, the $R^2$ of the Risk model (13.8\%) was higher than that of the Movies model (11.4\%).

While our level of precision was relatively low and direct pricing according to the prediction might not be profitable due to noise,\footnote{For example, when the model's predictive power is low, the high price might be shown too often to those whose WTP is above the anonymous price but below the high price, thereby resulting in no sale under personalized pricing. Figure \ref{fig:predictionPlots} in Appendix \ref{section:AppendixA} presents a plot of elicited versus predicted WTP for the training data.} the pricing model could help to identify participants with high and low WTP. As we are interested in pricing models with the same scope of prices between the two treatments, we opted for a pricing that offers three price levels: low, medium, and high. We fixed the medium price to be the anonymous price---that is, £1.85---and ran simulations for the high and low prices to maximize the profit for the Risk and the Movies surveys, such that the expected sales are higher than those under anonymous pricing. More precisely, we varied the high and low prices and the cutoff values that separate these prices in each model. The resulting pricing models  given below:

\begin{itemize}
  \item The Risk treatment: If the predicted WTP according to the Risk model is above £2.11, display the price £2.09. If the predicted WTP according to the Risk model is below £1.30, display the price £1.09. Otherwise, display £1.85.
  \item The Movies treatment: If the predicted WTP according to the Movies model is above £2.22, display the price £2.09. If the predicted WTP according to the Movies model is below £1.30, display the price £1.09. Otherwise, display £1.85.
\end{itemize}

It is worth noting that the objective of our model was not to approximate firms' more sophisticated pricing models in reality but rather to allow for some price discrimination in the experiment. The current model fits this purpose. We were primarily interested in participants' strategic responses to the announced use of their answers and their privacy choices. As we intentionally did not reveal any details of the model to the participants, we believe our results regarding treatment differences are independent of the exact model we use.\footnote{The results in  Section \ref{sec:buyingDecisions} are the exception, as indicated at the beginning of that section. However, these are not essential for the paper's main objective.}

The resulting simulated sales were £973.48 in the Risk survey and £981.08 in the Movies survey. While the pricing model does not offer a vast improvement over the anonymous prices, it suits our goal of testing the strategic response of the participants when facing personalized pricing models, which is the main focus of this paper.

\subsection{Strategic responses in the surveys}

We began by identifying attempts to strategize the survey responses by comparing the responses in the treatments to those in the training data. We sequentially analyze the Risk and the Movies treatments. 

\begin{figure}[htp]
  \centering
  \includegraphics[width=0.7\textwidth]{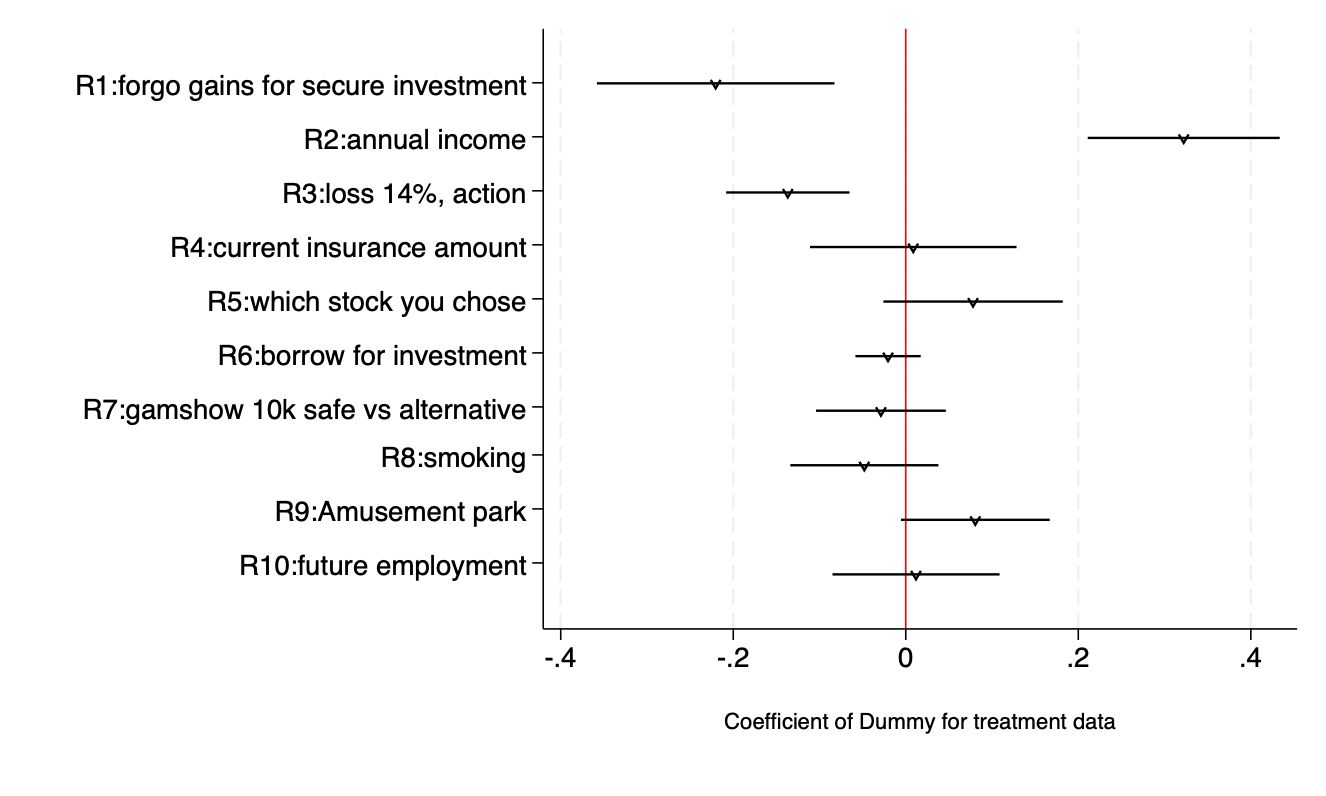}
  \caption{Coefficients of the Risk treatment dummy in OLS regressions versus training data}
  \label{fig:diffStrategiesRisk}
  \medskip 
  
\begin{minipage}{1\textwidth}
\footnotesize
Notes: The sample consists of the Risk treatment and the training data. A positive coefficient indicates that the average survey answer is larger than the training data. The questions are coded such, that higher values are associated with higher risk-taking. For example, for question R1, a response of strongly disagree with the statement corresponds to 5, while strongly agree  corresponds to 1. For question R2, a response of the annual income grows substantially corresponds to 5, while a substantial decrease corresponds to 1. For question R3, a response of buying more the investment when it loses 14\% of its value corresponds to 3, while selling the investment corresponds to 1.  
\end{minipage}
\end{figure}

First, we ran OLS regressions of the answers to each Risk survey question on a binary dummy variable to indicate whether the subject belongs to the Risk treatment or the training data. The risk dummy variable was assigned a value of 1 for participants in the Risk treatment and 0 otherwise. Age and gender were included as control variables in these regressions.
Figure \ref{fig:diffStrategiesRisk} visualizes the risk dummy coefficients for responses to the Risk survey questions among subjects in the Risk treatment and the training data. 
Relative to the training data, the average disagreement with the statement of forgoing gains in exchange for security (R1: Forgo gains for secure investment) by subjects in the treatment was significantly lower. 
In addition,  subjects in the Risk treatment significantly more often report that they expect a substantial annual income increase (R2: Annual income) and intend to  ``sell not to worry'' in the case of a sharp loss of 14\% of investment value (R3: Loss of 14\%, action). The evidence above strongly supports the manipulation hypothesis, as moving the average responses requires coordinated efforts to bias answers in a particular direction. 

\begin{table}[htp]\centering
\caption{$P$-values of the variance and Mann-Whitney tests for equality of answers in the training data and the Risk survey}\label{tab:varranksumStand}
\footnotesize
\begin{tabular}{l|r|r}\toprule
Question & $p$-value variance test& $p$-value Mann-Whitney test\\\midrule
R1: Forgo gains for secure investment &0.00&0.02 \\
R2: Annual income&0.00&0.00  \\
R3: Loss of 14\%, action &0.31&0.00 \\
R4: Current insurance amount &0.08&0.71 \\
R5: Which stock you choose &0.62&0.03 \\
R6: Borrow for investment &0.00&0.15 \\
R7: Gameshow 10k safe vs alternative &0.82&0.48 \\
R8: Smoking &0.00&0.19 \\
R9: Amusement park &0.30&0.00\\
R10: Future employments &0.69&0.32 \\
\bottomrule
\end{tabular}
\begin{minipage}{0.85\textwidth}
%Notes: * $p<0.10$, ** $p<0.05$, *** $p<0.01$.
\end{minipage}
\end{table}

We noted that while the differences in the responses to questions R1 and R3 align with the expectation for a manipulation aimed at lowering the lottery prices, the manipulations we infer on question R2 appear to go against that intuition and the correlations we observed with the WTP. The reasons for these discrepancies are unclear. One possibility is that participants in the Risk treatment may be signaling a strong desire to increase their earnings, thus potentially influencing the lottery price they need to pay. 

Another sign of manipulation attempts could be differences in the variance of answers or in the distributions. We test the former with a variance test and the latter with a Mann-Whitney non-parametric test. Table \ref{tab:varranksumStand} presents the $p$-values for the test of equality of variance (the second column) and equality of distributions (the third column) between the answers in the training data and the treatments. 
There is a significantly higher variance in the treatment than in the training data in the following questions: R1: Forgo gains for secure investment, R6: Borrow for investment, and R8: Smoking. There is a significantly lower variance in the treatment than in the training data in R2:Annual income. The results from the Mann-Whitney test generally align with the regression results, although they add significant differences in distributions in questions R5:Which stock you choose and R9:Amusement park. Thus, overall, we see at least one sign of manipulation (5\% significant difference in mean, variance, or distribution) in seven out of ten Risk survey questions. 

\begin{figure}[htp]
  \centering
  \includegraphics[width=0.7\textwidth]{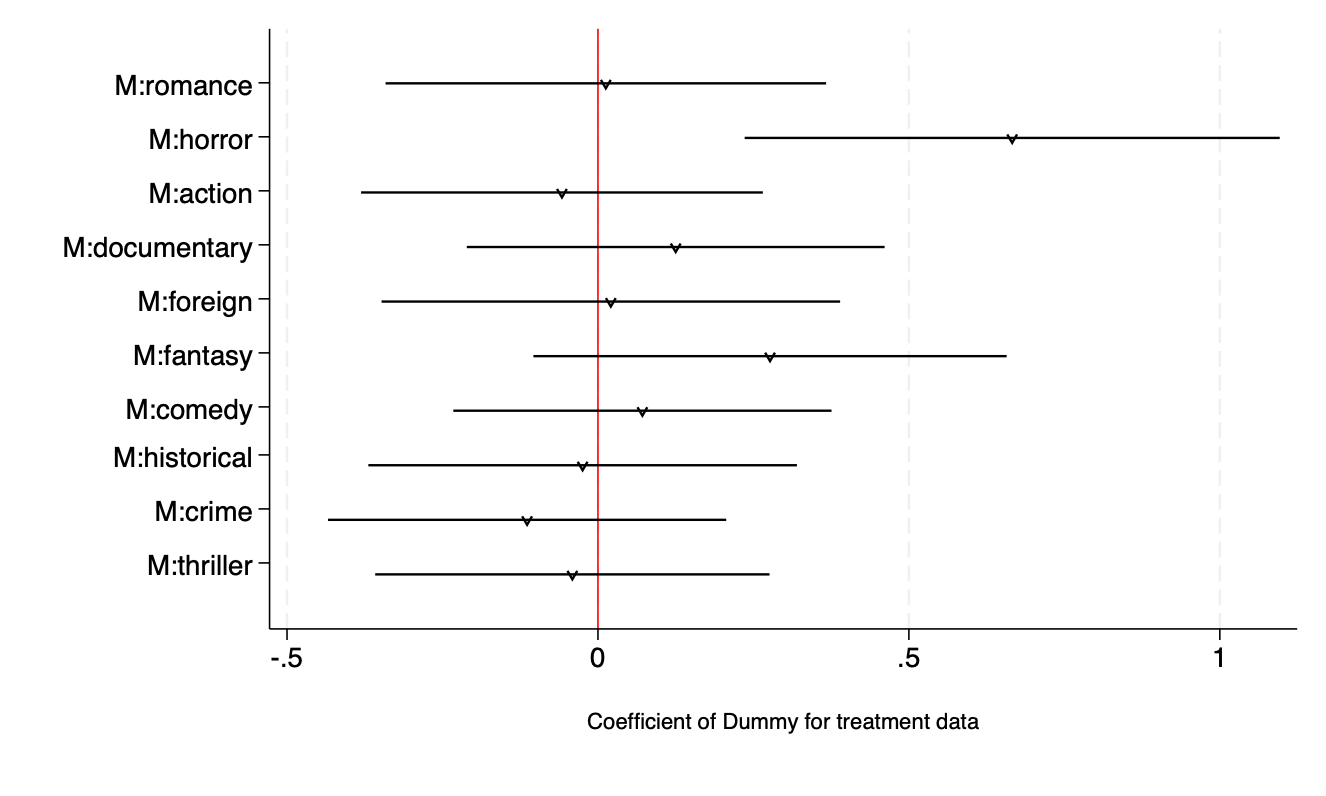}
  \caption{Coefficients of the Movies treatment dummy in OLS regressions versus the training data}
  \label{fig:diffStrategiesMovies}
\begin{minipage}{1\textwidth}
\medskip 

\footnotesize
Notes: The sample consists of the Movie treatment and the training data. OLS regressions control for gender, age, and age squared. A positive coefficient indicates that the average survey answer is larger than the training data. 
\end{minipage}
\end{figure}

In the Movies treatment, the only significant difference between the treatment and the training data in the OLS regressions is the higher ratings of horror movies in the treatment, as depicted in Figure \ref{fig:diffStrategiesMovies} (M2:horror). As earlier, the regressions include control for gender and age. 
There is a significantly higher variance in the treatment than in the training data in the rating of action movies. There was a significantly lower variance in the treatment than in the training data in the ratings of fantasy movies. In Table \ref{tab:varranksumMovies}, results from the Mann-Whitney test generally confirm the regression results for ratings of horror movies and also indicate a significantly different distribution of the ratings of fantasy movies between the training sample and the treatments. Thus, overall, there was at least one sign of manipulation (5\% significant difference in mean, variance, or distribution) in two out of ten Movies survey questions. An alternative interpretation is that, in the Movies survey, subjects manipulated their responses but in random directions. We cannot exclude this possibility, but it would also be a sign of a lack of agreement on how movie ratings relate to prices.

\begin{table}[htp]
\caption{$P$-values of variance and Mann-Whitney tests for equality of answers in the training data and treatments in the Movies survey}\label{tab:varranksumMovies}
\centering 
\begin{threeparttable}
    \footnotesize
\begin{tabular}{l|r|r}\toprule
Question &$p$-value variance test& $p$-value Mann-Whitney test\\\midrule
M1: Romance &0.42&0.80 \\
M2: Horror&0.97&0.00  \\
M3: Action &0.09&0.97 \\
M4: Documentary &0.51&0.98 \\
M5: Foreign&0.37&0.94 \\
M6: Fantasy &0.05&0.00 \\
M7: Comedy &0.22&0.18 \\
M8: Historical&0.86&0.07 \\
M9: Crime &0.97&0.42\\
M10: Thriller&0.85&0.82 \\
\bottomrule
\end{tabular}
\end{threeparttable}
\end{table}

\begin{innercustomres}
\textbf{(Strategic responses)}: Subjects attempted to manipulate their responses to the survey questions. We found significant manipulation in seven questions in the Risk survey and two in the Movies survey.
\end{innercustomres} 

Are these manipulations successful? In other words, do participants manage to lower the predicted WTP? 
Table \ref{tab:hyp2} presents the regression results that compare the predicted WTP  between the training data and treatments. Predicted WTP in the Risk treatment (Columns (1) and (2)) is significantly lower than in the training data. Thus, on average, participants' strategic responses are successful, thereby resulting in lower WTP estimates. Our previous result reveals an unexpected manipulation in question R2, which is contrary to expectations. Despite this apparently irrational strategy, participants still managed to influence the predicted WTP. This outcome occurs because the final model used to predict WTP does not include the income expectation responses. These responses were statistically insignificant in predicting WTP when controlling for other variables.

However, in the Movies treatment (Columns (3) and (4)), the opposite is true: on average, predicted WTP was higher than in the training data. This increase is not significant once we control for age and gender. Thus, in the Movies treatment, participants could not shift the predicted WTP in their favor.

\begin{table}[!htp]
\caption{OLS regressions for testing Hypothesis 2}\label{tab:hyp2}
\centering
\begin{threeparttable}
    \footnotesize
\begin{tabular}{lrrrrrrr}\toprule
          &Predicted &Predicted &Predicted &Predicted &Individual &Individual \\
          &WTP &WTP &WTP &WTP &price &price \\
         &(1) &(2) &(3) &(4) &(5) &(6) \\\midrule
Risk   &      -0.081\sym{***}&   -0.088\sym{***}    &     &     &   &   \\
             & (0.024)   & (0.024)   &     &     &    &   \\
Movies   &     &     & 0.054\sym{**} & 0.021   & 0.085\sym{***}& 0.087\sym{***}\\
            &        &         & (0.023)     & (0.024)     & (0.016)     & (0.016)     \\
Age       &         &  -0.001     &         &  -0.003\sym{***}&         &  -0.001     \\
          &         & (0.001)     &         & (0.001)     &         & (0.001)     \\
Age squared            &                  &   -0.000\sym{*}  &                  &   -0.000\sym{***}&                  &    0.000\sym{*}  \\
                &                  &  (0.000)         &                  &  (0.000)         &                  &  (0.000)         \\

Female     &         &  -0.063\sym{***}&         &  0.001     &         &  -0.034\sym{**} \\
          &         & (0.022)     &         & (0.021)     &         & (0.016)     \\
Constant        &    2.166\sym{***}&    2.071\sym{***}&    2.177\sym{***}&    2.012\sym{***}&    1.926\sym{***}&    2.090\sym{***}\\
                &  (0.013)         &  (0.094)         &  (0.013)         &  (0.091)         &  (0.011)         &  (0.073)         \\
\hline
Observations  &   1033     &   1033     &   1024     &   1024     &   603     &   603     \\
Adjusted \(R^{2}\)&    0.010         &    0.017         &    0.004         &    0.024         &    0.046         &    0.051         \\
Sample &Train+Risk &Train+Risk &Train+Mov &Train+Mov &Risk+Mov &Risk+Mov \\
\bottomrule
\end{tabular}

\smallskip 

\begin{tablenotes}[flushleft]
\item \footnotesize Notes: OLS regressions of the predicted WTP in Columns (1)---(4). OLS regressions of individual prices based on survey answers in Columns (5) and (6). Standard errors are given between parentheses. \sym{*} \(p<0.10\), \sym{**} \(p<0.05\), \sym{***} \(p<0.01\)\\
\end{tablenotes}
\end{threeparttable}
\end{table}

These first results of the data are informative, but does the shift in the predicted WTP have a meaningful impact on the price? Columns (5) and (6) in Table \ref{tab:hyp2} present the results of the regressions of the individual prices generated based on our pricing model and participants' responses. The individual prices are significantly higher in the Movies model than in the Risk model. These results directly support Hypothesis 2.

\begin{innercustomres}
\textbf{(Success of strategic response)}: The predicted WTP is significantly lower in the Risk survey than in the training data. There is no significant difference in the predicted WTP between the Movies and the training data, when controlling for age and gender. Individual prices are significantly higher in the Movies survey than in the Risk survey.
\end{innercustomres} 

\subsubsection{Robustness of manipulations}
\label{subsec:robustnessAndData}

One potential concern is that informing respondents regarding the upcoming lottery sale might bias their responses. Consequently, the observed disparities between the training data and the treatments could stem from the announcement of the possibility of buying a lottery itself rather than from a strategic attempt to influence the price of the lottery.
To mitigate this concern, we conducted two control treatments, exogenous price risk (ExogRisk) and exogenous price movies (ExogMovies). In these treatments, participants were explicitly informed that the lottery price would be independent of their responses:

\textit{``After these 10 questions, we will offer you an option to buy a lottery ticket from us, which gives you a 50\% chance of winning £5.
Note that the survey and the lottery are not connected; the survey is simply used to provide you with sufficient cash to potentially buy the lottery. The price of the lottery ticket is predetermined."}

The price offered to participants was consistently £1.85, which corresponded to the anonymous price in the Risk and Movies treatments. The treatments were run in Prolific in April 2024 (approximately 18 months later than the original sessions). We used a between-subjects design, with approximately 150 participants in each treatment.

We replicated the analyses above, focusing on the differences in responses between the Risk and Movies treatments relative to the ExogRisk and ExogMovies treatments, respectively.

\begin{figure}[htp]
  \centering
  \includegraphics[width=1\textwidth]{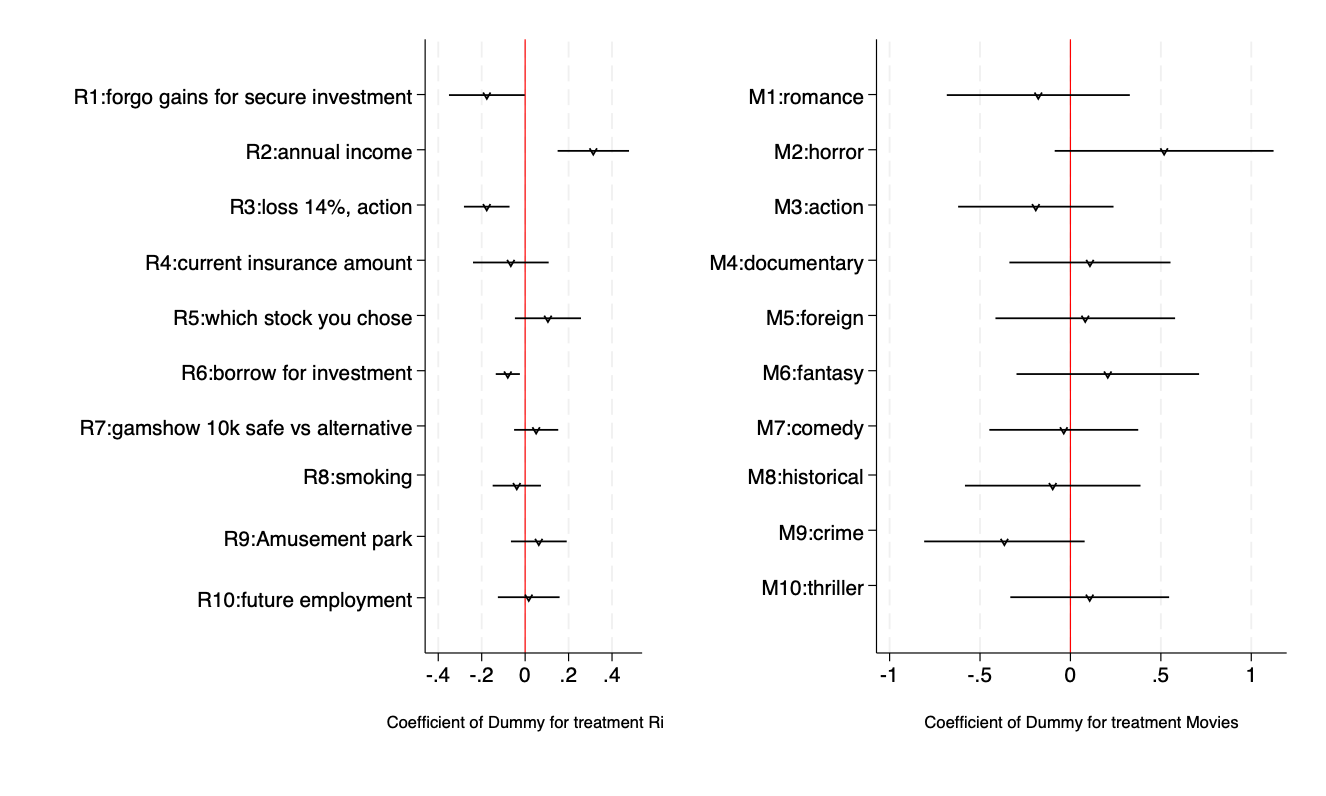}
  \caption{Coefficients of the main treatment dummy in OLS regressions versus exogenous price treatments. }
  \label{fig:robgraph}
\begin{minipage}{1\textwidth}
\medskip 

\footnotesize
Notes: On the left panel, the sample consists of the Risk treatment and the ExogRisk data. On the right panel, the sample consists of the Movies treatment and the ExogMovies data. OLS regressions control for gender, age, and age squared.  A positive coefficient indicates that the average survey answer in the Risk and Movies treatments is larger than in the ExogRisk and ExogMovies treatments, respectively. 
\end{minipage}
\end{figure}

Figure \ref{fig:robgraph} displays the coefficients of the differences between the main treatments and the treatments with the exogenous price. The left panel compares Risk and ExogRisk. Analogous to the comparison between Risk and the training data, subjects significantly more often report that they expect a substantial annual income increase (R2:Annual income), and intend to ``sell not to worry" in the case of a sharp loss of 14\% of investment value (R3:Loss of 14\%, action) in Risk than in ExogRisk. However, the average disagreement with the statement of forgoing gains in exchange for security (R1:Forgo gains for secure investment) by subjects in Risk is only marginally significantly higher than that in ExogRisk. Additionally, a significantly higher proportion of subjects report the experience of borrowing for investment (R6:Borrow for investment) in Risk than in ExogRisk, while we observed only differences in variance and not in the mean with respect to training data.

With regard to the Movies survey, Figure \ref{fig:robgraph} shows there are no significant differences between ratings in ExogMovies and Movies. Note that this is inconsistent with the results of the comparison between Movies and training data, where the average rating of horror movies differed significantly. In case of ExogMovies, the difference goes in the same direction as in training data, but it is not significant, potentially due to a smaller sample. However, note that Mann-Whitney test marginally rejects the equality of distributions of horror movies ratings between Movies and ExogMovies, as is evident from Table \ref{tab:varranksumMoviesrob}.

Tables \ref{tab:varranksumStandrob} and \ref{tab:varranksumMoviesrob} display the $p$-values for the tests of equality of variance (in the second column) and equality of distributions (in the third column) between the responses in the treatments with exogenous price and the main treatments.

\begin{table}[htp]\centering
\caption{$P$-values of variance and Mann-Whitney tests for equality of answers in the ExogRisk and the Risk survey}\label{tab:varranksumStandrob}
\footnotesize
\begin{tabular}{l|r|r}\toprule
Question & $p$-value variance test& $p$-value Mann-Whitney test\\\midrule
R1: Forgo gains for secure investment &0.09&0.046 \\
R2: Annual income&0.00&0.00  \\
R3: Loss of 14\%, action &0.64&0.00 \\
R4: Current insurance amount &0.18&0.63 \\
R5: Which stock you choose &0.65&0.18 \\
R6: Borrow for investment &0.00&0.00 \\
R7: Gameshow 10k safe vs alternative &0.08&0.50 \\
R8: Smoking &0.41&0.40 \\
R9: Amusement park &0.84&0.29\\
R10: Future employments &0.74&0.91 \\
\bottomrule
\end{tabular}
\begin{minipage}{0.85\textwidth}
\end{minipage}
\end{table}

\begin{table}[htp]
\caption{$P$-values of variance and Mann-Whitney tests for equality of answers in the training data and treatments in the Movies survey}\label{tab:varranksumMoviesrob}
\centering 
\begin{threeparttable}
    \footnotesize
\begin{tabular}{l|r|r}\toprule
Question &$p$-value variance test& $p$-value Mann-Whitney test\\\midrule
M1: Romance &0.54&0.55 \\
M2: Horror&0.31&0.09  \\
M3: Action &0.64&0.34 \\
M4: Documentary &0.09&0.38 \\
M5: Foreign&0.68&0.72 \\
M6: Fantasy &0.86&0.35 \\
M7: Comedy &0.89&0.85 \\
M8: Historical&0.51&0.74 \\
M9: Crime &0.64&0.11\\
M10: Thriller&0.27&0.38 \\
\bottomrule
\end{tabular}
\end{threeparttable}
\end{table}

Comparing Risk with ExogRisk, we observe signs of manipulation in four out of ten questions at the 5\% significance level (five out of ten if we count the marginally significant results at the 10\% level). When comparing Movies with ExogMovies, we find signs of manipulation in zero out of ten questions (one out of ten if we count the marginally significant results at the 10\% level).

Table \ref{tab:hyp2rob} in the Appendix presents an analog of Table \ref{tab:hyp2}. The regression results indicate that the predicted WTP for the lottery is significantly lower in Risk than in ExogRisk, while there is no significant difference in the predicted WTP between Movies and ExogMovies.

Thus, overall, we conclude that the results are robust to the announcement of the lottery and are driven by the attempts of participants to manipulate the price of the lottery in the Risk treatment.

The ExogRisk and ExogMovies treatments also allow us to test whether the fact that our training sample is representative of the U.S. population (while the treatment samples are not) could explain a few of our results.

Table \ref{tab:demographics} reveals a substantial difference in age between the training and other samples. One could question if the effects that we report between the training and treatment samples could be explained by differences in the sample populations. However, throughout our analysis, we included controls for age and gender in our regression analyses. The results ensure that the observed differences in outcomes are attributable to the treatment effects rather than underlying demographic variations.

Moreover, the ExogRisk and ExogMovies treatments closely matched the demographic characteristics of our original non-training samples, particularly in terms of gender balance and age distribution. The consistency of the results further reinforces the reliability of our findings.

\begin{table}[ht]
\centering
\caption{Summary statistics for demographic variables across different samples}
\label{tab:demographics}
\begin{threeparttable}
\begin{tabular}{@{}lccccc@{}}
\toprule
& Training & Risk & Movies & ExogRisk & ExogMovies \\ \midrule
Age & 44.2 (15.79) & 35.96 (12.37) & 35.44 (11.22) & 37.05 (13.15) & 35.88 (11.95) \\
Female  & 0.507 (0.500) & 0.500 (0.501) & 0.502 (0.501) & 0.493 (0.502) & 0.493 (0.502) \\
\bottomrule
\end{tabular}
\medskip
\begin{tablenotes}[flushleft]
\item \small Notes: A t-test of equality of average age reveals a significant difference when comparing the training group to all other treatment groups. All other pairwise comparisons are not significant (minimum $p$-value is 0.18).
\end{tablenotes}
\end{threeparttable}
\end{table}

\subsection{Privacy choices}

Before analyzing the treatment differences, we first look at the correlates of the privacy choices in our baseline treatments of the Risk and Movies surveys. Table \ref{tab:descrpriv} presents the results of the marginal effects of a Probit regression for the decision to choose the privacy option or not on the observables. Column (1) shows that the privacy choice does not correlate with age and gender. 
Column (2) controls for the beliefs regarding the highest and lowest possible individual prices.\footnote{The average beliefs for the lower bound of the pricing model were $£1.03$ in the Risk treatment and $£1.06$ in the Movies treatment. The average beliefs for the upper bound of the pricing model were $£2.58$ in the Risk treatment and $£2.76$ in the Movies treatment. Therefore, in both cases, their average beliefs were in line with the pricing model results, which could be above or below the anonymous price.}  When participants have higher beliefs regarding the upper and lower bounds of individual prices, they are less likely to choose the privacy option. 
However, the coefficients are only marginally significant. The interpretation is challenging, as one would expect the opposite direction of the effects. In other words, we should expect that the higher the beliefs regarding the price bounds, the more likely participants are to choose the privacy option, as it allows them to hide information and avoid personalized pricing. 

One possible explanation for the negative correlation between privacy choices and beliefs regarding the highest and lowest prices   
is that participants base their privacy choices on their expectations of the anonymous price. Even though they believe the low and high prices are high, they may still choose to disclose their information if they expect the alternative anonymous price to also be high. Unfortunately, we did not elicit this in our study. Nevertheless, if participants believed that the anonymous price would fall between the highest and the lowest prices, then we can use the average between the lowest and highest beliefs as an approximation of the believed anonymous price. Column (3) controls for this average belief. The coefficient is negative and significant: the larger the average belief, the less often participants choose the privacy option. 
This is in line with the interpretation that a higher believed anonymous price leads to lower privacy demand. 

\begin{table}[!htp]
\centering
\caption{Correlates of privacy choices in baselines (Risk and Movies) }\label{tab:descrpriv}
\begin{threeparttable}
    \footnotesize 
\begin{tabular}{lrrrr}\toprule
        &Privacy choice&Privacy choice&Privacy choice\\
        &(1)&(2)&(3)\\\midrule

Age             &   -0.003         &   -0.006         &   -0.006         \\
                &  (0.008)         &  (0.008)         &  (0.008)         \\
Age squared           &    0.000         &    0.000         &    0.000         \\
                &  (0.000)         &  (0.000)         &  (0.000)         \\
Female          &    0.044         &    0.031         &    0.032         \\
                &  (0.036)         &  (0.037)         &  (0.036)         \\
Belief low        &                  &   -0.082\sym{*}  &                  \\
                &                  &  (0.042)         &                  \\
Belief high      &                  &   -0.052\sym{*}  &                  \\
                &                  &  (0.031)         &                  \\
Average belief    &                  &                  &   -0.127\sym{***}\\
                &                  &                  &  (0.045)         \\

\hline
Observations    &      603         &      602         &      602         \\
\bottomrule
\end{tabular} 
\begin{tablenotes}[flushleft]
\item Notes: The table reports the marginal effect of the Probit regression of a dummy variable for the choice of privacy option. The sample includes both Risk and Movies treatments. Belief low indicates the elicited belief regarding the lowest possible individual price. Belief high is the elicited belief regarding the highest possible individual price. The average belief reveals the average between believed low and high individual prices. Standard errors are given between parentheses. \sym{*} \(p<0.10\), \sym{**} \(p<0.05\), \sym{***} \(p<0.01\)\\
\end{tablenotes}
\end{threeparttable}
\end{table}

Next, we analyze treatment differences in the privacy choices.
Overall, we see low demand for privacy (31.1\% and 23.2\% in Risk and Movies surveys, respectively). Table \ref{tab:Hyps3and5} presents the marginal effects of a Probit regression on privacy choices. Column (1) indicates that participants chose the privacy option significantly less often in Movies than in Risk.  The result is robust to individual controls and a proxy for beliefs regarding the anonymous price, as shown in Columns (2) and (3). Thus, we find no support for Hypothesis 3 but a significant opposite effect. 

We propose three possible explanations for this result. The first possibility is participants' overconfidence in their ability to manipulate the Movies survey.\footnote{Overconfidence has been found in different applications, including lying \citep{serra2021mistakes}, unraveling of the matching market \citep{dargnies2019self}, labor market \citep{santos2020overconfidence}, delegation to algorithms \citep{dargnies2022algoaversion} and market places like that in our paper \citep{grubb2015overconfident}.} The link between the context of the survey and the product is weak in this survey. Therefore, the pricing model is less clear to participants compared to that in the Risk survey. However, their attempts to game the system, which in fact reduced their welfare, are consistent with the fact that they overestimated their ability to understand the pricing model.

The second possibility is that participants perceive the Movies survey as less informative to the firm for the purpose of price discrimination. In fact, this perception is corroborated by the low correlation between the responses to that survey and WTP as well as the lower $R^2$ in the estimated pricing model. Therefore, the lower demand for the privacy option in the Movies survey could reflect a rational response to their understanding of the limited pricing relevance of their movie preferences: if responses to the Movies survey do not impact prices, one could expect that any response would result, perhaps on average, in the same price as the one obtained with the privacy option. 
While we cannot distinguish between these explanations, we believe the latter is more realistic:  the seemingly unrelated context of the movie ratings leads to subjects believing that it is less consequential for the price.

Finally, the third explanation could be that subjects perceive the information in the Risk treatment as more ``sensitive''---in the sense of revealing information that one would not like others to have access to---than in the Movies treatment. This perception requires a degree of naïvete, as privacy, in the context of our survey, merely implies that the answers are not used in pricing but are still available to researchers. In our robustness experiments (ExofRisk and ExogMovies), we asked subjects to rate how sensitive the information in the survey was on a scale from 0 to 10. The average sensitivity rating was 3.51 for the Risk survey and 1.57 for the Movies survey ($p<0.01$). While the general average sensitivity is low, the responses in the Risk survey are perceived as more sensitive. This might drive the result of lower demand for privacy if subjects predominantly perceive the option of buying privacy as a means to conceal their answers to the survey.\footnote{In addition to these questions, we also tested whether subjects were simply giving the same response to the survey questions, as a means to avoid providing useful information, by calculating the standard deviation for the responses provided by each individual. These show that even among those with the lowest value, the standard deviation value was approximately 0.50, thereby leading us to reject that hypothesis. More details can be found in Tables \ref{tab:RiskSurveyresultsLowestStd} and \ref{tab:MoviesSurveyresultsLowestStd} in the Appendix. We thank an anonymous referee for suggesting this potential explanation.}

\begin{table}[htp]
\centering
\caption{Treatment difference in privacy choices }\label{tab:Hyps3and5}
\begin{threeparttable}
    \footnotesize 
\begin{tabular}{lrrrr}
\toprule
      &Privacy choice&Privacy choice&Privacy choice\\
        &(1)&(2)&(3)\\\midrule

Movies             &   -0.078\sym{**} &   -0.078\sym{**} &   -0.074\sym{**} \\
                &  (0.036)         &  (0.036)         &  (0.036)         \\
Age             &                  &   -0.002         &   -0.004         \\
                &                  &  (0.008)         &  (0.008)         \\
Age squared            &                  &    0.000         &    0.000         \\
                &                  &  (0.000)         &  (0.000)         \\
Female          &                  &    0.045         &    0.032         \\
                &                  &  (0.036)         &  (0.036)         \\
Average belief    &                  &                  &   -0.125\sym{***}\\
                &                  &                  &  (0.045)         \\
\hline
Observations    &      603         &      603         &      602         \\
\bottomrule
\end{tabular}
\begin{tablenotes}[flushleft]
 \item Notes: The table reports marginal effects of Probit regressions on the choice of the privacy option dummy. The sample includes both Risk and Movies treatments. The average belief reveals the average between the believed low and high individual prices. Standard errors are given in parentheses. \sym{*} \(p<0.10\), \sym{**} \(p<0.05\), \sym{***} \(p<0.01\)\\
\end{tablenotes}
\end{threeparttable}
\end{table}

Now, we proceed to analyze the optimality of the privacy choices. We constructed a dummy variable for optimal privacy choice, which is equal to 1 if: (i) the individual price is high, and the participant chooses the privacy option; (ii) the individual price is low or middle, and the participant chooses not to buy the privacy option; it equals zero otherwise.\footnote{Our analysis assumes that everyone prefers to see a lower price. This is not exactly optimal since when a participant does not buy the lottery, we do not know the conterfactual decision. Thus, optimality analysis is impossible in the strict sense. Nevertheless, we think the current approach is informative about optimal sorting into privacy, at least to some extent.}

\begin{figure}
  \centering
  \includegraphics[width=0.46\textwidth]{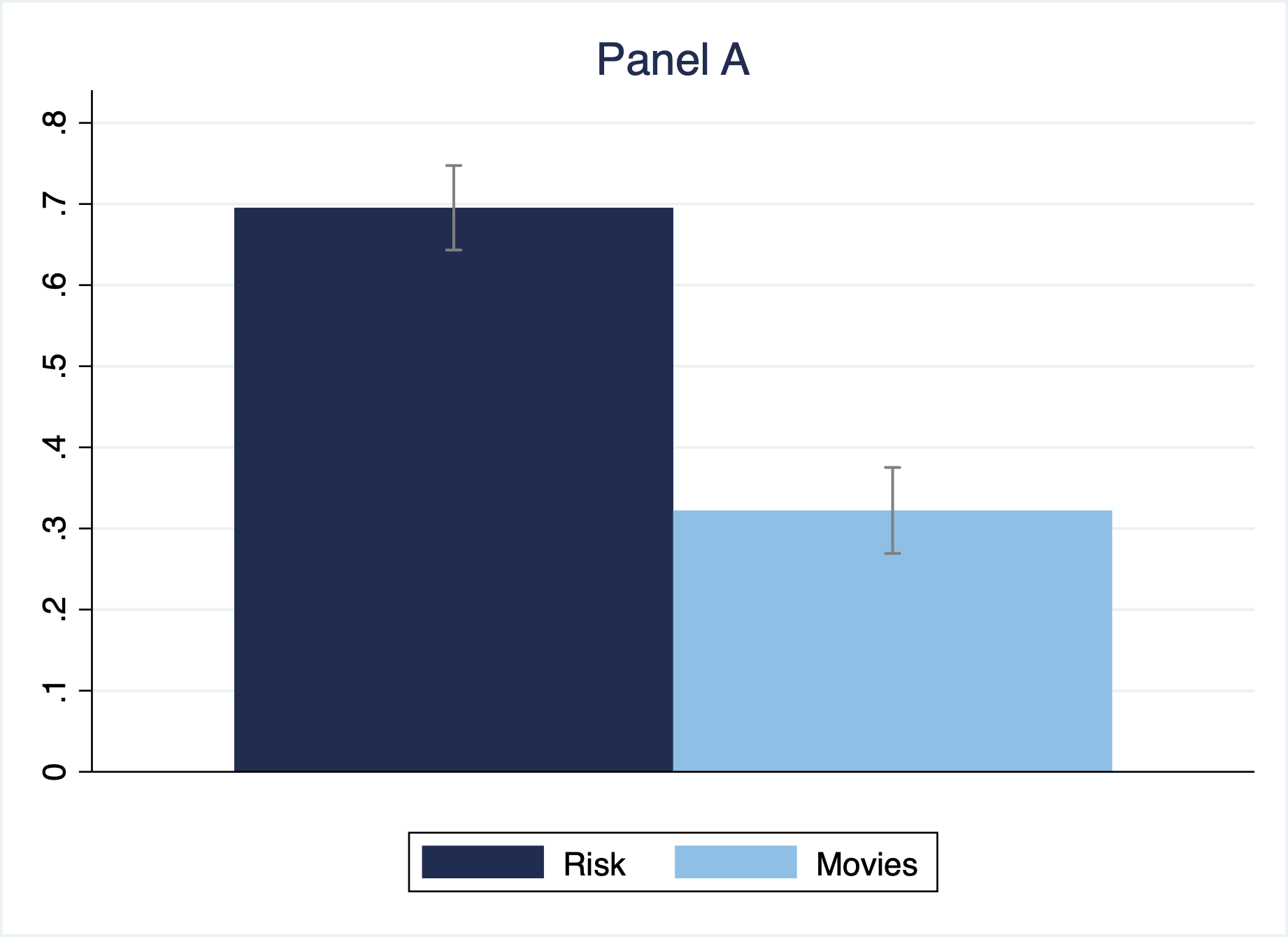}\includegraphics[width=0.46\textwidth]{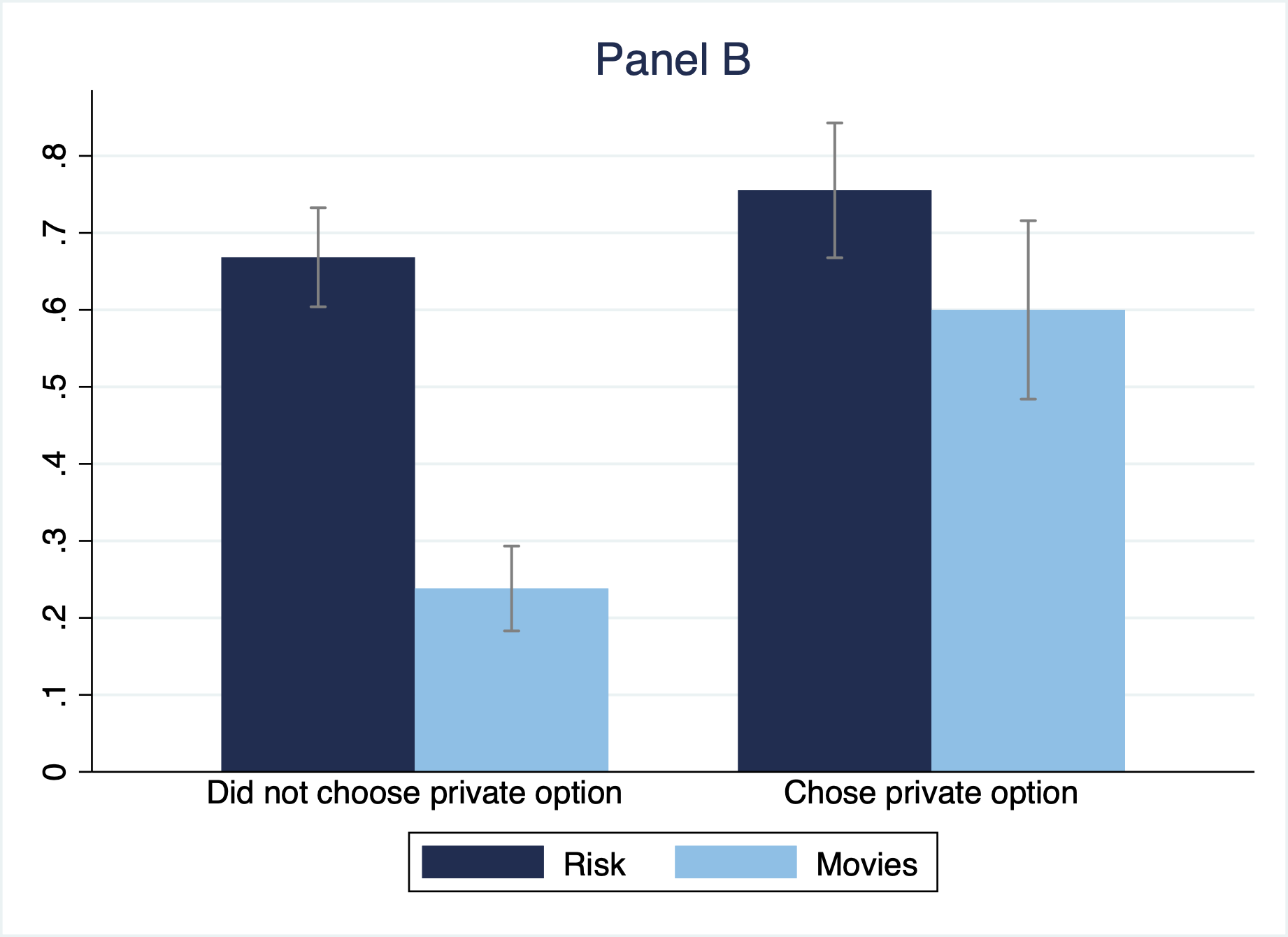}
  \caption{Proportions of optimal choices of privacy}
  \label{fig:optimalChoicePrivacy}
\end{figure}

Panel A of Figure \ref{fig:optimalChoicePrivacy} depicts the proportion of optimal privacy choices by treatment. The proportion is higher in the Risk survey than in the Movies survey, and the difference is large. Table \ref{tab:Hyps4and6} presents the results of regression analyses, and Columns (1) and (2) support the significance of the difference. This supports Hypothesis 4. Thus, participants are better sorted in selecting the privacy option in the Risk survey as compared to in the Movies survey, which suggests that participants have a better understanding of the relationship between the responses and the resulting prices in the Risk survey as compared to that in the Movies survey. 

Panel B of Figure \ref{fig:optimalChoicePrivacy} presents the proportion of optimal privacy choices by treatment, dividing the sample according to whether participants chose the privacy option. The figure suggests that the main difference in the optimality of the privacy choice comes from those who did not choose the privacy option. In the Movies survey, only 24\% of participants who did not choose the privacy option did it optimally, while this proportion is 67\% in the Risk survey. Columns (3) and (4) of Table \ref{tab:Hyps4and6} present the regression analyses, controlling for privacy choice. First, on average, the proportion of optimal choices is higher among those who chose the privacy option (see the coefficient of the variable ``Chose privacy dummy" in Column (3)). However, this result is driven by the Movies treatment, as seen in Column (4). Thus, the inferior proportion of optimal privacy choices in the Movies survey is driven by the under-demand for privacy from those who would face the high price. This result is in line with the explanation that the participants accurately believe that the Movies survey was less \textbf{informative} to the firm for the purpose of price discrimination, but misunderstood that for meaning that prices would not be \textbf{sensitive} to their responses. Importantly, treatment differences are significant, even for those who chose the privacy option ($p=0.04$).

\begin{table}[!htp]
\centering
\caption{Treatment difference in the optimality of privacy choice}\label{tab:Hyps4and6}
\footnotesize
\begin{threeparttable}

    \begin{tabular}{lrrrr}\toprule
&Optimal privacy  &Optimal privacy &Optimal privacy &Optimal privacy  \\
& choice &choice & choice  &choice \\
 &(1)&(2)&(3)&(4)\\\midrule

Movies             &   -0.345\sym{***}&   -0.344\sym{***}&   -0.327\sym{***}&   -0.387\sym{***}\\
                &  (0.029)         &  (0.029)         &  (0.029)         &  (0.033)         \\
Age             &                  &    0.001         &    0.001         &    0.001         \\
                &                  &  (0.002)         &  (0.002)         &  (0.002)         \\
Female          &                  &    0.056         &    0.039         &    0.039         \\
                &                  &  (0.038)         &  (0.037)         &  (0.037)         \\
Chose privacy dummy         &                  &                  &    0.202\sym{***}&    0.083         \\
                &                  &                  &  (0.040)         &  (0.056)         \\
Movies*Chose privacy dummy        &                  &                  &                  &    0.240\sym{***}\\
                &                  &                  &                  &  (0.080)         \\
\hline
Observations    &      603         &      603         &      603         &      603         \\
\bottomrule
\end{tabular}
\begin{tablenotes}[flushleft]
\item Notes: The table reports the marginal effects of Probit regressions of dummy for optimal choice of privacy option. The sample includes all treatments. Standard errors are given in parentheses. \sym{*} \(p<0.10\), \sym{**} \(p<0.05\), \sym{***} \(p<0.01\)\\
\end{tablenotes}
\end{threeparttable}
\end{table}

\begin{innercustomres}
\textbf{(Privacy choices)}: The proportion of participants who chose the privacy option in the Risk survey is significantly higher than that in the Movies survey. The proportion of optimal decisions of whether to buy the privacy option is significantly higher in the Risk survey than in the Movies survey. The largest treatment difference in optimality is driven by those who did not buy the privacy option.
\end{innercustomres}

\subsection{Buying decisions and payoffs}
\label{sec:buyingDecisions}

In a last step, we examine results for the buying decisions and payoffs. These results are presented for the completeness of the analysis. It should be taken with a grain of salt, as the buying decisions and payoffs rely heavily on the precision of the algorithm that predicts the WTP and, thus, may not be externally valid, unlike the main results of the paper regarding the strategic responses and privacy choices. 

Table \ref{tab:buying} presents the marginal effects of a Probit regression for the dummy of buying the lottery. Columns (1) and (2) show that the proportion of buying decisions is significantly lower in the Movies survey than in the Risk survey. However, controlling for the price, there is no significant difference between the two treatments, as seen in Column (3). The higher the price, the lower the propensity to buy. 
Column (4) adds a control for those who chose the privacy option and suggests that those who opted for privacy are significantly more likely to buy the lottery. One explanation could be that this is the effect of price. Column (5) restricts the sample only to those who saw the same price, £1.85, which is the anonymous price and the price for those predicted to have middle valuations in our pricing models. The dummy for choosing the privacy option remains significant. Thus, the effect is not driven by price. It is either the consequence of sorting, with those who have higher valuations buying the privacy option more often, or the behavioral effect of ``safety" to buy under anonymous prices, without the perception that one might be tricked into buying for a high individual price.

\begin{table}[!htp]
\centering
\caption{Buying decisions}\label{tab:buying}
\begin{threeparttable}
    \scriptsize
\begin{tabular}{lrrrrr}\toprule
&Bought lottery &Bought lottery&Bought lottery&Bought lottery &Bought lottery \\
&(1) &(2)&(3)&(4)&(5) \\\midrule

Movies           &   -0.098\sym{**} &   -0.096\sym{**} &   -0.045         &   -0.044         &   -0.060         \\
                &  (0.040)         &  (0.041)         &  (0.042)         &  (0.042)         &  (0.056)         \\
Age             &                  &   -0.009         &   -0.012         &   -0.012         &   -0.007         \\
                &                  &  (0.010)         &  (0.009)         &  (0.009)         &  (0.012)         \\
Age squared            &                  &    0.000         &    0.000         &    0.000         &    0.000         \\
                &                  &  (0.000)         &  (0.000)         &  (0.000)         &  (0.000)         \\
Female          &                  &    0.026         &    0.009         &    0.007         &   -0.048         \\
                &                  &  (0.041)         &  (0.041)         &  (0.040)         &  (0.053)         \\
Price    &                  &                  &   -0.480\sym{***}&   -0.393\sym{***}&                  \\
                &                  &                  &  (0.124)         &  (0.126)         &                  \\
Chose privacy dummy         &                  &                  &                  &    0.104\sym{**} &    0.113\sym{**} \\
                &                  &                  &                  &  (0.046)         &  (0.053)         \\
\hline
Observations    &      603         &      603         &      603         &      603         &      342         \\
\hline
Sample  &   All     &   All     &   All  &   All    &   Price = anonymous     \\
\bottomrule
\end{tabular}
\begin{tablenotes}[flushleft]
\item Notes: The table reports the marginal effects of  Probit regressions of a dummy for buying the lottery. The sample in Columns (1)---(4) includes all data from treatments. The sample in Column (5) includes only those who saw the anonymous price, either because they chose the privacy option or because their predicted WTP suggested the middle price in the pricing model.  Standard errors are given in parentheses. \sym{*} \(p<0.10\), \sym{**} \(p<0.05\), \sym{***} \(p<0.01\)\\
\end{tablenotes}
\end{threeparttable}
\end{table}

Next, we examine at the payoffs, which serve as a proxy for consumer welfare. We consider payoffs only for the main binary decisions in the experiment---that is, whether to buy privacy and whether to buy the lottery. To calculate the payoffs, we first assign a payoff of £0 to all participants and then add  £2.50 minus the price for those who bought the lottery. We use  £2.50 as the expected lottery payoff to avoid noise due to the randomization of the lottery payoff. We deduct £0.10 for those who chose privacy. The resulting values represent the payoffs. 

\begin{figure}[htp]
  \centering
  \includegraphics[width=0.6\textwidth]{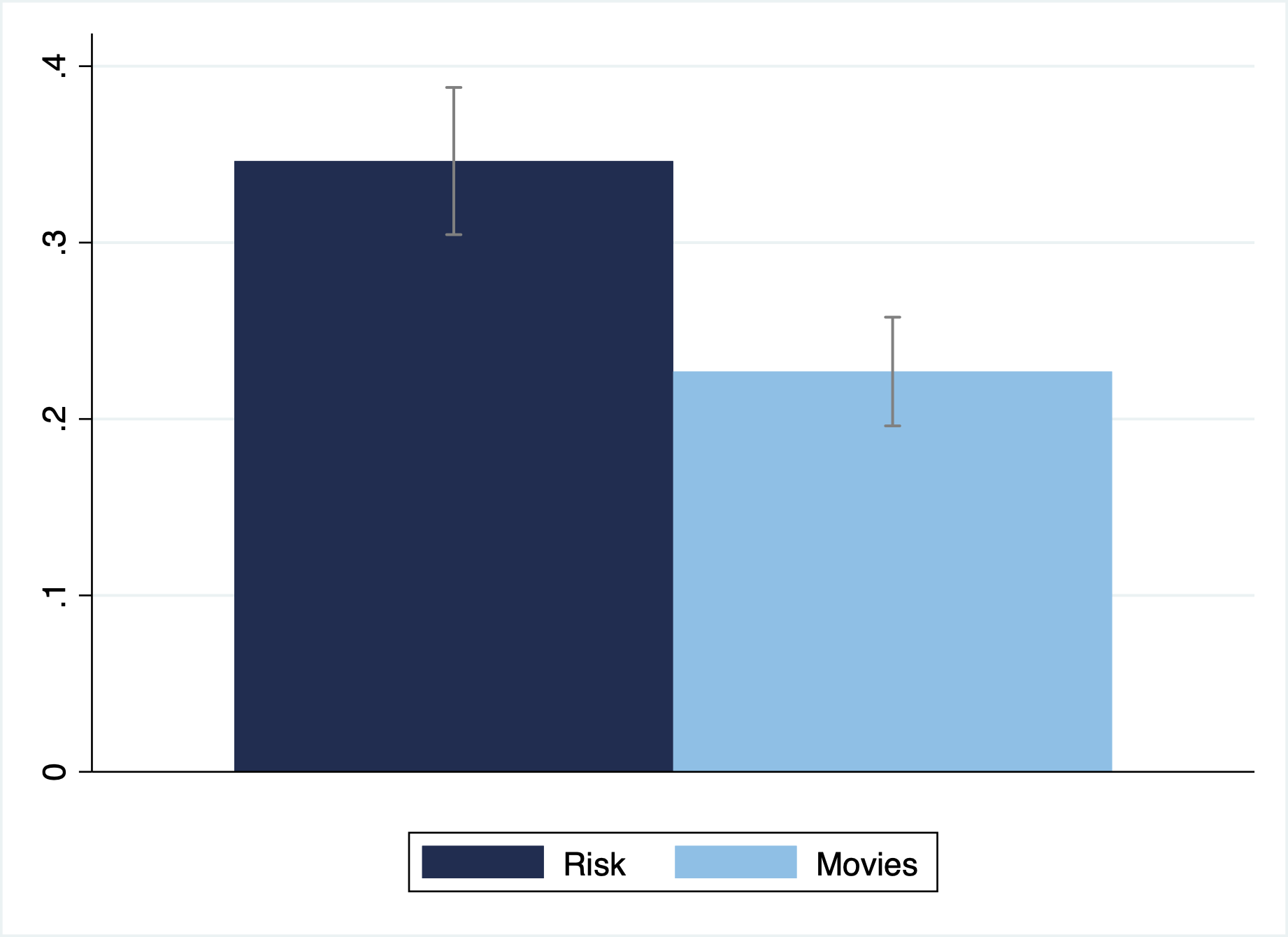}
  \caption{Payoffs by treatment}
  \label{fig:payoffs}
\end{figure}

Figure \ref{fig:payoffs} displays the average payoffs by treatment, and Table \ref{tab:payoff} details the results of the OLS regression of participants' payoff for the main task, with Column (1) reporting that the payoff is significantly lower in Movies than in Risk. Further, the significance of the difference in payoffs is robust to individual controls, as seen in Column (2). Thus, participants earn significantly less in Movies than in Risk, as they are less able to respond strategically to the firm's screening. Column (3) provides additional evidence, thereby revealing that the treatment difference goes completely through the differences in prices. 

\begin{table}[!htp]
\centering
\caption{Payoffs of participants}\label{tab:payoff}
\footnotesize 
\begin{threeparttable}
    \begin{tabular}{lrrr}\toprule
&Payoff &Payoff&Payoff\\
&(1) &(2)&(3)\\\midrule
Movies             &   -0.119\sym{***}&   -0.120\sym{***}&   -0.025         \\
                &  (0.026)         &  (0.026)         &  (0.025)         \\
Age             &                  &    0.002         &   -0.004         \\
                &                  &  (0.006)         &  (0.006)         \\
Age squared            &                  &   -0.000         &    0.000         \\
                &                  &  (0.000)         &  (0.000)         \\
Female          &                  &    0.033         &   -0.001         \\
                &                  &  (0.027)         &  (0.024)         \\
Price    &                  &                  &   -0.852\sym{***}\\
                &                  &                  &  (0.069)         \\
Constant        &    0.346\sym{***}&    0.291\sym{**} &    2.023\sym{***}\\
                &  (0.019)         &  (0.124)         &  (0.179)         \\
\hline
Observations    &      603         &      603         &      603         \\
Adjusted \(R^{2}\)&    0.031         &    0.029         &    0.224          \\
\hline
Sample  &   All     &   All     &   All        \\
\bottomrule
\end{tabular}
\begin{tablenotes}[flushleft]
\item Notes: OLS regression of participants' payoff for the main task.  Standard errors are given in parentheses. \sym{*} \(p<0.10\), \sym{**} \(p<0.05\), \sym{***} \(p<0.01\)\\
\end{tablenotes}
\end{threeparttable}
\end{table}

Finally, we also examine revenues and find no significant treatment differences (see Table \ref{tab:revenue} in Appendix \ref{section:AppendixA}. 

\begin{innercustomres}
\textbf{(Buying decisions and payoffs)}: There is no significant difference in the proportion of participants who buy the lottery between treatments, controlling for the prices. The payoffs of participants are significantly higher in the Risk survey than in the Movies survey.
\end{innercustomres} 
 
\section{Discussion and conclusions}\label{section:conclusion}

In this paper, we experimentally studied consumers' strategic responses to personalized pricing by exploiting the link between their behavior and the pricing model. Our results provide a novel angle on analyzing consumers' decisions regarding big data. The ability of consumers to effectively manage their privacy depends on their understanding of how their data is used for pricing. 
Participants in our main experiment stage were incentivized to manipulate their responses to surveys. We distinguished between two ways of consumer profiling. One, based on the methods before big data became readily available, is when a firm determines prices based on consumer responses to a survey in the same context as the product and, thus, has a more explicit link between answers to its questions and the resulting prices. The other, based on the possibility of exploiting sophisticated statistical relations with big data, is when a firm determines prices based on consumer answers to a survey in a context that is different from the product. As expected, they were more successful in manipulation when the context is similar. This suggests an additional ``vulnerability'' for consumers in the era of big data, as it is more difficult to gain from personalized pricing through strategic responses. We also conjectured that participants were more likely to pay for privacy when the link was less clear, as they should anticipate the difficulty of manipulating. Our result suggests the opposite: participants in the survey of a seemingly unrelated context demanded less privacy, and fewer decisions were optimal than those in a closer context. This surprising result was driven by those who did not buy privacy, even though they should have. 

How externally valid are our results? Our experiments presented an artificial and rather simplified setup to study price discrimination, and our personalized price models are clearly less precise than those used by firms in the real world. However, we believe this fact does not affect the interpretations regarding the main interest of our paper---that is the attempts and the ability to strategically manipulate responses. Our results in the Movies survey, in which participants failed to strategically respond to personalized pricing successfully, are likely to overestimate the degree of strategic response because we presented the best conditions for it. Moreover, in terms of privacy, we believe that our treatment differences are independent of the model's precision and reflect participants' perception of when they are more likely to benefit from private browsing. Furthermore, using a lottery as the product for sale in our setup raised the concern that risk preferences could correlate with demand for privacy. Although testing this correlation proves challenging due to the manipulated responses in our treatments, it is worth noting that even if this correlation exists, it does not undermine the validity of our results when comparing between treatments.  Nonetheless, exploring this correlation in future research would provide valuable insights into how risk preferences affect privacy demand beyond the strategic context studied in this paper. 

One important policy implication of our study is that with the increasing use of consumer information in seemingly unrelated contexts for pricing strategies, consumer protection becomes even more important than before.  When firms use traditional personalized pricing, certain consumers might have reasonably simple strategies to recover some of the welfare that firms capture with price discrimination. When these pricing models use more sophisticated relations and seemingly unrelated variables, questions such as the extent to which consumers understand these relations and their own ability to manipulate them might be of significance. The lack of clarity not only hinders  consumers' ability to obtain desirable prices, but also reduces the volume of transactions, where misguided manipulations prevent sales from happening at all. 

Providing privacy options alone might not suffice to mitigate this problem. While current data protection regulations place greater emphasis on transparency in the data collection, future policies should also encourage transparency in pricing algorithms and providing consumers with better information on how their data is used. This could further improve consumers' decisions while still leaving space for welfare-improving price discrimination. Returning to the personalized pricing practices mentioned in the Introduction, we are witnessing some exciting developments in policy to address this issue. The Dutch Authority for Consumers and Markets recently made a landmark decision to order Wish, an online shopping platform, to be transparent regarding their algorithm. 

Another implication from our results is that the prominence of the privacy option could be context-dependent. For example, instead of bombarding cyberspace with cookie choices in every interaction, these could become more prominent in situations in which consumers are identified with significant vulnerabilities for the suboptimal use of cookies. Another related open question is whether an opt-in system in which the default option of privacy protection and consumers choose for personalized pricing if they want, instead of an opt-out system like we studied in this paper, could reduce strategic mistakes. 

Finally, our results suggest that using observed privacy choices from the field for welfare analysis may be problematic. When consumers do not choose privacy options, one might conclude that there is a lack of demand. However, it might instead be that consumers are not making informed choices, and the demand would change if they were better informed. 

Taken together, our results indicate that policies that promote consumer awareness of data transparency complement regulations that require firms to abide by data transparency and provide consumers with control choices. While recent public debates have mostly centered around the regulation approach of ``notice and consent,'' it is also essential to educate consumers. One without the other could harm consumer welfare.

\bibliographystyle{plainnat}
\bibliography{priceDisc}

\pagebreak{}

\appendix

\section*{Appendix}

\setcounter{table}{0}
\renewcommand{\thetable}{A.\arabic{table}}

\setcounter{figure}{0}
\renewcommand{\thefigure}{A.\arabic{figure}}

\section{Additional tables and figures\label{section:AppendixA}}

\begin{table}[h!]
\centering
\caption{OLS regression of female dummy in the Movies survey}\label{tab:AppFemalemov}
\begin{threeparttable}
    \footnotesize
{
\def\sym#1{\ifmmode^{#1}\else\(^{#1}\)\fi}
\begin{tabular}{l*{1}{c}}
\hline\hline
                &\multicolumn{1}{c}{Female}\\
\hline
M1: Romance              &    0.077\sym{***}\\
                &  (0.007)         \\
M2: Horror              &    0.006         \\
                &  (0.006)         \\
M3: Action              &   -0.056\sym{***}\\
                &  (0.009)         \\
M4: Documentary              &    0.030\sym{***}\\
                &  (0.008)         \\
M5: Foreign              &   -0.023\sym{***}\\
                &  (0.007)         \\
M6: Fantasy              &    0.000         \\
                &  (0.007)         \\
M7: Comedy              &   -0.021\sym{**} \\
                &  (0.008)         \\
M8: Historical             &   -0.017\sym{**} \\
                &  (0.008)         \\
M9: Crime             &   -0.003         \\
                &  (0.010)         \\
M10: Thriller             &   -0.008         \\
                &  (0.010)         \\
Constant        &    0.661\sym{***}\\
                &  (0.083)         \\
\hline
Observations    &      723         \\
\(R^{2}\)       &    0.248         \\
Adjusted \(R^{2}\)&    0.237         \\
sample     &    Training     \\
\hline\hline

\end{tabular}
}
\begin{tablenotes}[flushleft]
\item 
Notes: OLS regression of a female dummy on the ratings on the Movies survey.  The female dummy is assigned a value of 1 for female subjects and 0 for male subjects. Standard errors are given in parentheses. \sym{*} \(p<0.10\), \sym{**} \(p<0.05\), \sym{***} \(p<0.01\)\\
\end{tablenotes}
\end{threeparttable}
\end{table}

\begin{table}[h!]\centering
\caption{Results of the linear discriminant analysis for predicting gender from the Movies survey}\label{tab:AppFemalemovDisc}
\footnotesize
{
\def\sym#1{\ifmmode^{#1}\else\(^{#1}\)\fi}
\begin{tabular}{l|c|c}
\hline\hline
                &Classified female & Classified male\\
\hline
Female             &    275& 85 \\
                &  76.4\% & 23.6\%        \\
                \hline
Male              &    102& 261 \\
                &  28.1\% & 71.9\%        \\

\hline\hline

\end{tabular}
}

\end{table}

\clearpage

\begin{table}[h!]
\centering
\caption{OLS regression of WTP in the Risk survey}\label{tab:Appwtpstand}
\footnotesize
\begin{threeparttable}
    {
\def\sym#1{\ifmmode^{#1}\else\(^{#1}\)\fi}
\begin{tabular}{l*{1}{c}}
\hline\hline
        &\multicolumn{1}{c}{WTP}\\
\hline
R1: Forgo gains for secure investment&  -1.450\sym{***}\\
        & (0.213)     \\
R7: Gameshow 10k safe vs alternative&  0.146\sym{**} \\
        & (0.064)     \\
R1 $\times$ R1     &  0.172\sym{***}\\
        & (0.029)     \\
R1 $\times$ R4     &  0.085\sym{***}\\
        & (0.024)     \\
R3 $\times$ R1    &  0.057\sym{**} \\
        & (0.023)     \\
R4 $\times$ R6     &  -0.095\sym{**} \\
        & (0.047)     \\
R8 $\times$ R1     &  0.159\sym{***}\\
        & (0.039)     \\
R8 $\times$ R9     &  -0.160\sym{***}\\
        & (0.047)     \\
R9 $\times$ R6     &  0.244\sym{***}\\
        & (0.065)     \\
Constant    &  3.036\sym{***}\\
        & (0.302)     \\
\hline
Observations  &   731     \\
\(R^{2}\)    &  0.138     \\
Adjusted \(R^{2}\)&  0.127     \\
sample     &    Training     \\
\hline\hline

\end{tabular}
}

\begin{tablenotes}[flushleft]
\item Notes: OLS regression of the WTP on answers to the Risk survey. R1 $\times$ R1 is the squared of the answer R1: Forgo gains for secure investment. R1 $\times$ R4 is the interaction between R1: Forgo gains for secure investment and R4: Current insurance amount. R3 $\times$ R1 is the interaction between R3: Loss of 14\% and R1: Forgo gains for secure investment. R4 $\times$ R6 is the interaction between R4: Current insurance amount and R6: Borrow for investment. R8 $\times$ R1 is the interaction between R8: Smoking and R1: Forgo gains for secure investment. R8 $\times$ R9 is the interaction between R8: Smoking and R9: Amusement park. R9 $\times$ R6 is the interaction between R9: Amusement park and R6: Borrow for investment. Standard errors are given in parentheses. \sym{*} \(p<0.10\), \sym{**} \(p<0.05\), \sym{***} \(p<0.01\)\\
\end{tablenotes}
\end{threeparttable}
\end{table}

\begin{table}[h!]\centering
\caption{OLS regression of WTP in the Movies survey}\label{tab:Appwtpmov}
\footnotesize 
{
\def\sym#1{\ifmmode^{#1}\else\(^{#1}\)\fi}
\begin{tabular}{l*{1}{c}}
\hline\hline
        &\multicolumn{1}{c}{WTP}\\
\hline
M1: Romance   &  0.147\sym{**} \\
        & (0.069)     \\
M2: Horror    &  0.069\sym{*} \\
        & (0.042)     \\
M3: Action    &  -0.130\sym{***}\\
        & (0.035)     \\
M6: Fantasy   &  0.165\sym{***}\\
        & (0.049)     \\
M9: Crime    &  0.150\sym{***}\\
        & (0.030)     \\
M10: Thriller  &  -0.139\sym{***}\\
        & (0.041)     \\
M1 $\times$ M1     &  -0.011\sym{**} \\
        & (0.006)     \\
M1 $\times$ M2     &  -0.008\sym{*} \\
        & (0.004)     \\
M1 $\times$ M7     &  -0.014\sym{**} \\
        & (0.007)     \\
M2 $\times$ M8     &  0.011\sym{**} \\
        & (0.004)     \\
M3 $\times$ M1     &  0.019\sym{***}\\
        & (0.005)     \\
M4 $\times$ M2     &  -0.010\sym{*} \\
        & (0.005)     \\
M4 $\times$ M5     &  -0.014\sym{***}\\
        & (0.004)     \\
M4 $\times$ M6     &  -0.009\sym{*} \\
        & (0.005)     \\
M4 $\times$ M10    &  0.021\sym{***}\\
        & (0.005)     \\
M5 $\times$ M7     &  0.011\sym{***}\\
        & (0.004)     \\
M7 $\times$ M6     &  -0.014\sym{***}\\
        & (0.005)     \\
M7 $\times$ M7     &  0.011\sym{***}\\
        & (0.004)     \\
M8 $\times$ M9     &  -0.013\sym{***}\\
        & (0.004)     \\
Constant    &  1.371\sym{***}\\
        & (0.290)     \\
\hline
Observations  &   723     \\
\(R^{2}\)    &  0.114     \\
Adjusted \(R^{2}\)&  0.090    \\
sample     &    Training     \\
\hline\hline

\end{tabular}
}

\smallskip

\begin{minipage}{0.5\textwidth}
\footnotesize Notes: OLS regression of the WTP on answers to the Movies survey. MX $\times$ MY id interaction of MX and MY, where X and Y are between 1 and 10, and correspond to the index of question. M1: Romance, M2: Horror, M3: Action, M4: Documentary, M5: Foreign, M6: Fantasy, M7: Comedy, M8: Historical, M9: Crime, M10: Thriller. Standard errors are given in parentheses. \sym{*} \(p<0.10\), \sym{**} \(p<0.05\), \sym{***} \(p<0.01\)\\
\end{minipage}
\end{table}
\clearpage

\begin{figure}
    \centering
    \includegraphics[scale=0.3]{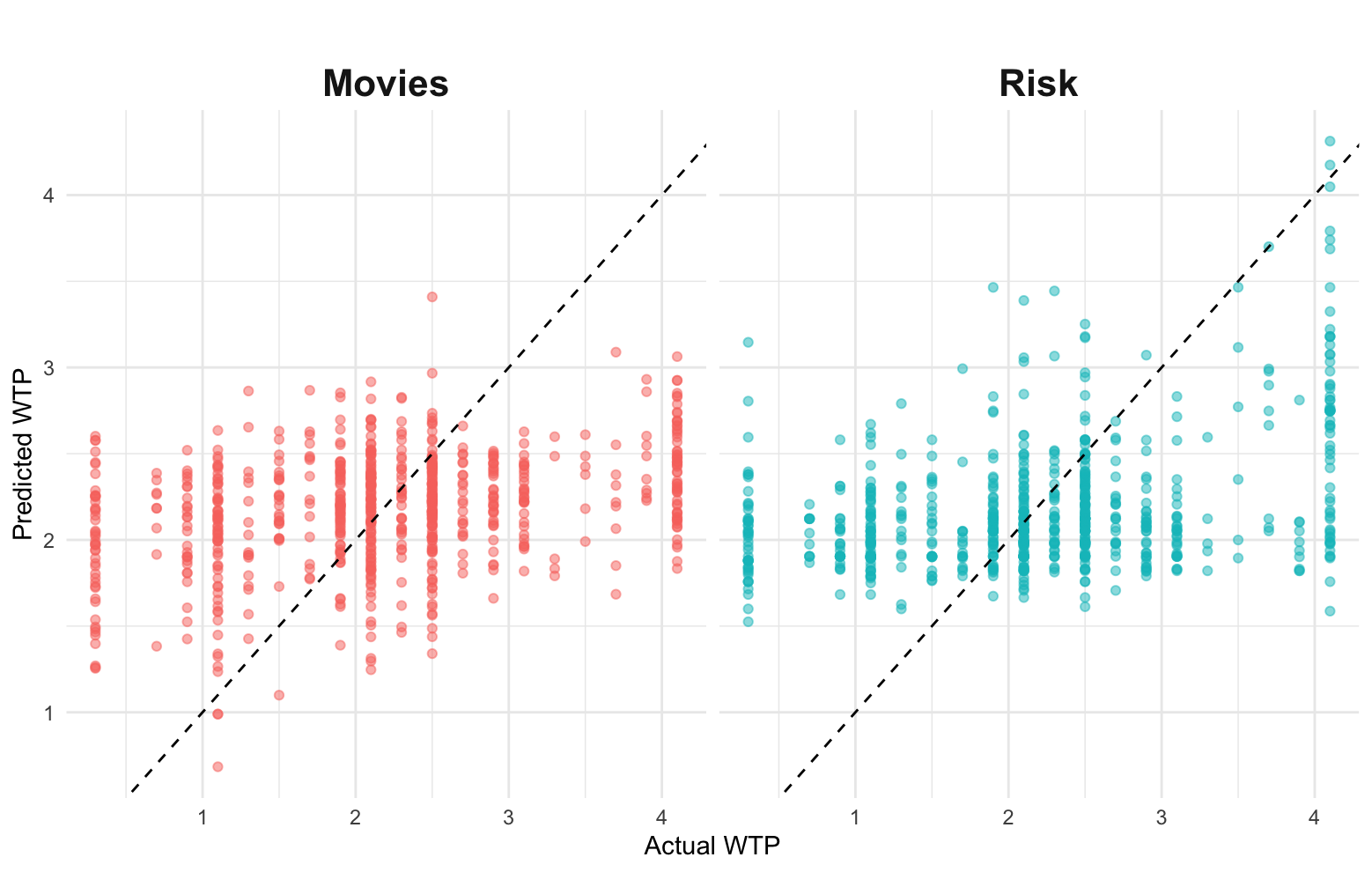}
    \caption{Predicted vs. Actual (Elicited) WTP for the training data}
    \label{fig:predictionPlots}
    \smallskip

\begin{minipage}{0.85\textwidth}
\footnotesize Note: For each subject in the training data, the value on the horizontal axis indicates the elicited WTP, and the one on the vertical axis the WTP predicted by the pricing model, given the answers of that subject to the related survey.
\end{minipage}
\end{figure}

\begin{table}[h!]
\centering
\caption{Risk survey responses with the lowest sstandard deviations} 
\label{tab:RiskSurveyresultsLowestStd}
\begin{tabular}{rrrrrrrrrrr}
  \toprule
StdDev & R1 & R2 & R3 & R4 & R5 & R6 & R7 & R8 & R9 & R10 \\ 
  \midrule
0.48 & 2.00 & 2.00 & 2.00 & 3.00 & 2.00 & 2.00 & 2.00 & 3.00 & 2.00 & 3.00 \\ 
  0.67 & 2.00 & 3.00 & 2.00 & 3.00 & 2.00 & 1.00 & 1.00 & 2.00 & 2.00 & 2.00 \\ 
  0.67 & 2.00 & 1.00 & 2.00 & 3.00 & 2.00 & 1.00 & 1.00 & 1.00 & 2.00 & 2.00 \\ 
  0.67 & 2.00 & 3.00 & 1.00 & 1.00 & 1.00 & 1.00 & 1.00 & 1.00 & 1.00 & 1.00 \\ 
  0.70 & 2.00 & 2.00 & 2.00 & 3.00 & 1.00 & 1.00 & 1.00 & 1.00 & 2.00 & 1.00 \\ 
   \bottomrule
\end{tabular}

\smallskip

\begin{minipage}{0.85\textwidth}
\footnotesize Note: The table presents the five survey responses with the lowest standard deviations across their responses to the ratings in the Movie survey. The columns represent the standard deviation and the answers to each of the 10 ratings.\\
\end{minipage}
\end{table}

\begin{table}[h!]
\centering
\caption{Movies survey responses with the lowest standard deviations} 
\label{tab:MoviesSurveyresultsLowestStd}
\begin{tabular}{rrrrrrrrrrr}
\toprule
StdDev & M1 & M2 & M3 & M4 & M5 & M6 & M7 & M8 & M9 & M10 \\ 
  \midrule
 0.57 & 7.00 & 6.00 & 6.50 & 8.00 & 7.50 & 7.00 & 7.50 & 7.00 & 7.00 & 7.50 \\ 
  0.68 & 5.50 & 5.50 & 6.50 & 5.50 & 6.00 & 7.00 & 7.50 & 6.50 & 6.00 & 6.50 \\ 
  0.70 & 6.50 & 8.00 & 8.00 & 8.00 & 6.00 & 7.50 & 8.00 & 7.00 & 7.50 & 7.50 \\ 
  0.74 & 7.50 & 6.00 & 5.50 & 6.00 & 7.50 & 7.50 & 7.00 & 6.50 & 6.00 & 6.50 \\ 
  0.75 & 9.00 & 7.50 & 9.00 & 10.00 & 9.00 & 9.00 & 10.00 & 10.00 & 9.00 & 9.00\\
  \bottomrule
  \end{tabular}

\smallskip

\begin{minipage}{0.85\textwidth}
\footnotesize Note: The table presents the five survey responses with the lowest standard deviations across their responses to the ratings in the Movie survey. The columns represent the standard deviation and the answers to each of the ten ratings.\\
\end{minipage}
\end{table}

\begin{table}[!htp]
\caption{Tests for the robustness of Hypothesis 2}\label{tab:hyp2rob}
\centering
\begin{threeparttable}
    \footnotesize
\begin{tabular}{lrrrrr}
\hline\hline
                &\multicolumn{1}{c}{Fitted values}&\multicolumn{1}{c}{Fitted values}&\multicolumn{1}{c}{Fitted values}&\multicolumn{1}{c}{Fitted values}\\
\hline
Risk          &   -0.062\sym{**} &   -0.061\sym{**} &                  &                  \\
                &  (0.031)         &  (0.030)         &                  &                  \\
 Movies        &                  &                  &   -0.015         &   -0.016         \\
                &                  &                  &  (0.033)         &  (0.033)         \\
Age             &                  &   -0.015\sym{**} &                  &    0.007         \\
                &                  &  (0.007)         &                  &  (0.008)         \\
Age squared            &                  &    0.000\sym{**} &                  &   -0.000         \\
                &                  &  (0.000)         &                  &  (0.000)         \\
female          &                  &   -0.124\sym{***}&                  &    0.006         \\
                &                  &  (0.029)         &                  &  (0.032)         \\

Constant        &    2.147\sym{***}&    2.506\sym{***}&    2.246\sym{***}&    2.139\sym{***}\\
                &  (0.025)         &  (0.135)         &  (0.027)         &  (0.152)         \\
\hline
Observations    &      452         &      452         &      453         &      453         \\
\hline
Observations    &      452         &      452         &      453         &      453         \\
Adjusted \(R^{2}\)&    0.007         &    0.038         &   -0.002         &   -0.001         \\
Sample &ExogRisk+Risk &ExogRisk+Risk &ExogMov+Mov &ExogMov+Mov  \\
\hline\hline

\end{tabular}

\smallskip 

\begin{tablenotes}[flushleft]
\item \footnotesize Notes: OLS regressions of the predicted WTP in Columns (1)---(4). OLS regressions of individual prices based on survey answers in Columns (5) and (6). Standard errors are given in parentheses. \sym{*} \(p<0.10\), \sym{**} \(p<0.05\), \sym{***} \(p<0.01\)\\
\end{tablenotes}
\end{threeparttable}
\end{table}

\begin{table}[h!]
\centering
\caption{Revenue}\label{tab:revenue}
\footnotesize
\begin{threeparttable}
    \begin{tabular}{lrr}\toprule
&Revenue &Revenue \\\midrule
Movies             &   -0.117         &   -0.119         \\
                &  (0.078)         &  (0.078)         \\
Age             &                  &   -0.002         \\
                &                  &  (0.003)         \\
Female          &                  &    0.031         \\
                &                  &  (0.078)         \\
Constant        &    1.046\sym{***}&    1.110\sym{***}\\
                &  (0.055)         &  (0.133)         \\
\hline
Observations    &      603         &      603         \\
Adjusted \(R^{2}\)       &    0.002         &    -0.002         \\
\bottomrule
\end{tabular}
\begin{tablenotes}[flushleft]
\item Notes: OLS of Revenues. Standard errors are given in parentheses. \sym{*} \(p<0.10\), \sym{**} \(p<0.05\), \sym{***} \(p<0.01\)
\end{tablenotes}
\end{threeparttable}
\end{table}

\newpage

\section{Instructions\label{section:Appendix_instructions}}

\subsection{Common to all treatments}

\textbf{Screen 1}

Consent:  You are invited to take part in a research study. The study is administered by researchers at the University of Lausanne, University of Gothenburg, and Southwestern University of Finance and Economics, in Chengdu.
\vspace{0.6cm}

You will receive £1 for participating. Total duration of the study is 5 to 6 minutes. 
\vspace{0.6cm}

All data will be treated confidentially. Data will be used anonymously and for academic research only. Anonymized data will be made available to other researchers for replication purposes. 

o	I understand the conditions and consent to participate in this study 

o	I reject participation

\vspace{0.6cm}

\textbf{Screen 2}

What is your gender?

o	Male  

o	Female  

o	Prefer not to answer  

\vspace{0.6cm}
Age: How old are you?

\vspace{0.6cm}

What is your Prolific ID? (Note that it should be filled automatically. If yes, just proceed further.)

\vspace{1.6cm}

\subsection{Training data}

\textbf{Screen 3}

In the next block, you will answer 21 questions about yourself. Please read the questions carefully and attempt to choose the answer that is as close to your preferences as possible. 

\vspace{0.6cm}
\textbf{Screens 4---24 Risk survey and Movies survey in random order}

\vspace{0.6cm}
\textbf{Risk Survey}

\vspace{0.6cm}
Q1.1 

I am prepared to forego potentially large gains if it means that the value of my investment is secure

o	I strongly agree  (1) 

o	I agree  (2) 

o	I neither agree or disagree  (3) 

o	I disagree  (4) 

o	I strongly disagree  (5) 

\vspace{0.6cm}

Q1.2 

Over the next several years, you expect your annual income to:

o	Stay about the same  (3) 

o	Grow moderately  (4) 

o	Grow substantially  (5) 

o	Decrease moderately  (2) 

o	Decrease substantially  (1)

 \vspace{0.6cm}

Q1.3

Imagine that due to a general market correction, one of your investments loses 14\% of its value a short time after you buy it. What do you do?

o	Sell the investment so you will not have to worry if it continues to decline  (1) 

o	Hold on to it and wait for it to climb back up  (2) 

o	Buy more of the same investment because at the current lower price, it looks even better than when you bought it  (3)

\vspace{0.6cm}
Q1.4 

What is the current amount of insurance you buy (life insurance, home insurance, medical insurance, travel insurance, etc)?

o	Much less than most of people I know  (5) 

o	Less than most people I know  (4) 

o	About the same as most people I know  (3) 

o	More than most people I know  (2) 

o	Much more than most people I know  (1)

\vspace{0.6cm}
Q1.5 

Assuming you are investing in a stock, which one would you choose?

o	Companies that may make significant technological advances that are still selling at their low initial offering price  (3) 

o	Established, well-known companies that have a potential for continued growth  (2) 

o	Established, stable, and well-recognized corporation that pay dividends  (1)

\vspace{0.6cm}
Q1.6 

Have you ever borrowed money for the purpose of making an investment (other than for marriage)?

o	Yes  (2) 

o	No  (1) 

\vspace{0.6cm}

Q1.7

You have just reached the \$10,000 plateau on a TV game show. Now you must choose between quitting with the \$10,000 in hand or betting the entire \$10,000 in one of three alternative scenarios. Which do you choose?

o	The \$10,000 -- you take the money and run  (1) 

o	A 50 percent chance of winning \$50,000  (2) 

o	A 20 percent chance of winning \$75,000  (3) 

o	A 5 percent chance of winning \$100,000  (4) 

\vspace{0.6cm}

Q1.8 

Do you smoke cigarettes?

o	Yes, daily  (3) 

o	Yes, occasionally  (2) 

o	No  (1) 

\vspace{0.6cm}

Q1.9 

In an amusement park, which of the following describes  your type best?

o	I always select the most extreme and exciting attractions, such as roller coasters with dead loops.  (3)

o	I look for enjoyable attractions with not too many extreme conditions.  (2) 

o	I prefer attractions with no adrenaline at all that offer quiet time and enjoyment of the atmosphere of the part, such as artistic performances.  (1)

\vspace{0.6cm}

Q1.10 Which of the following describes your preferences for future employment best?

o	I am self-employed and an owner of my business  (3) 

o	I work in a stable well paying government job  (1) 

o	I am a professional with stable income in a private firm  (2)

\vspace{0.6cm}
\clearpage

\textbf{Movies Survey}

\includegraphics[width=0.9\textwidth]{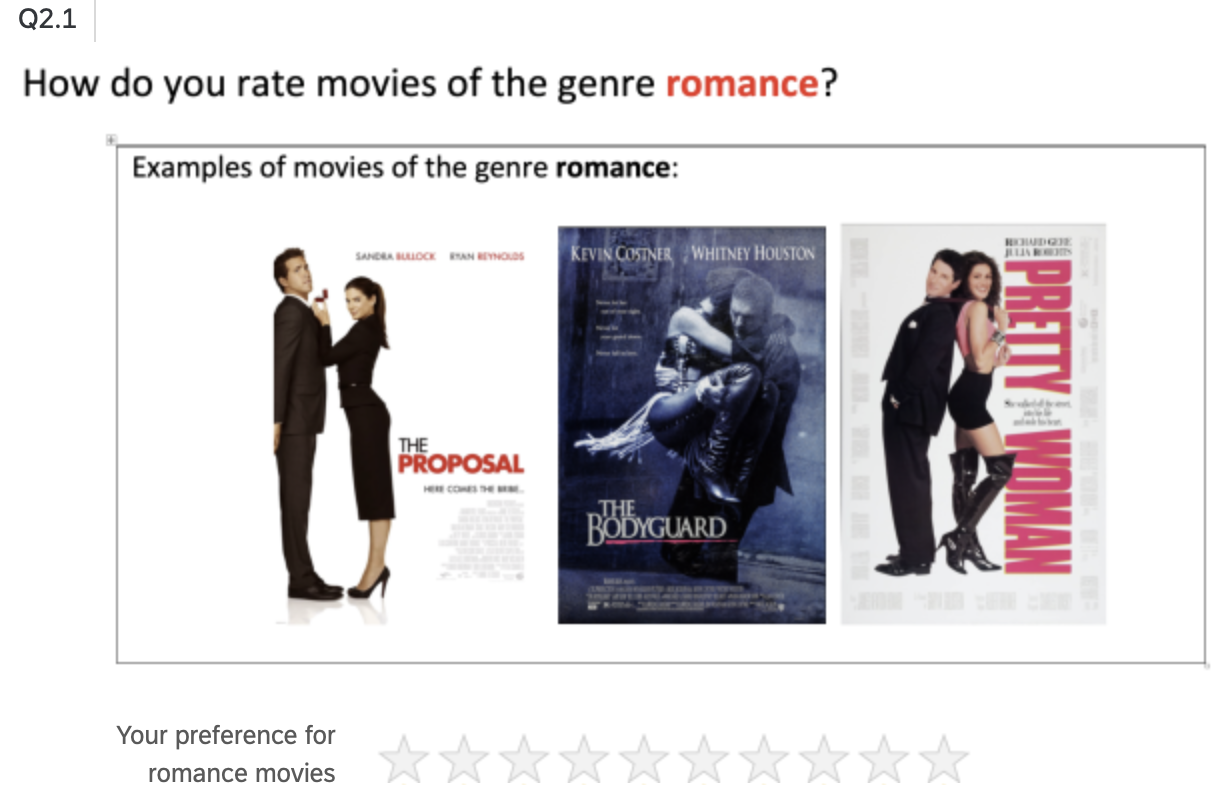}

\includegraphics[width=0.9\textwidth]{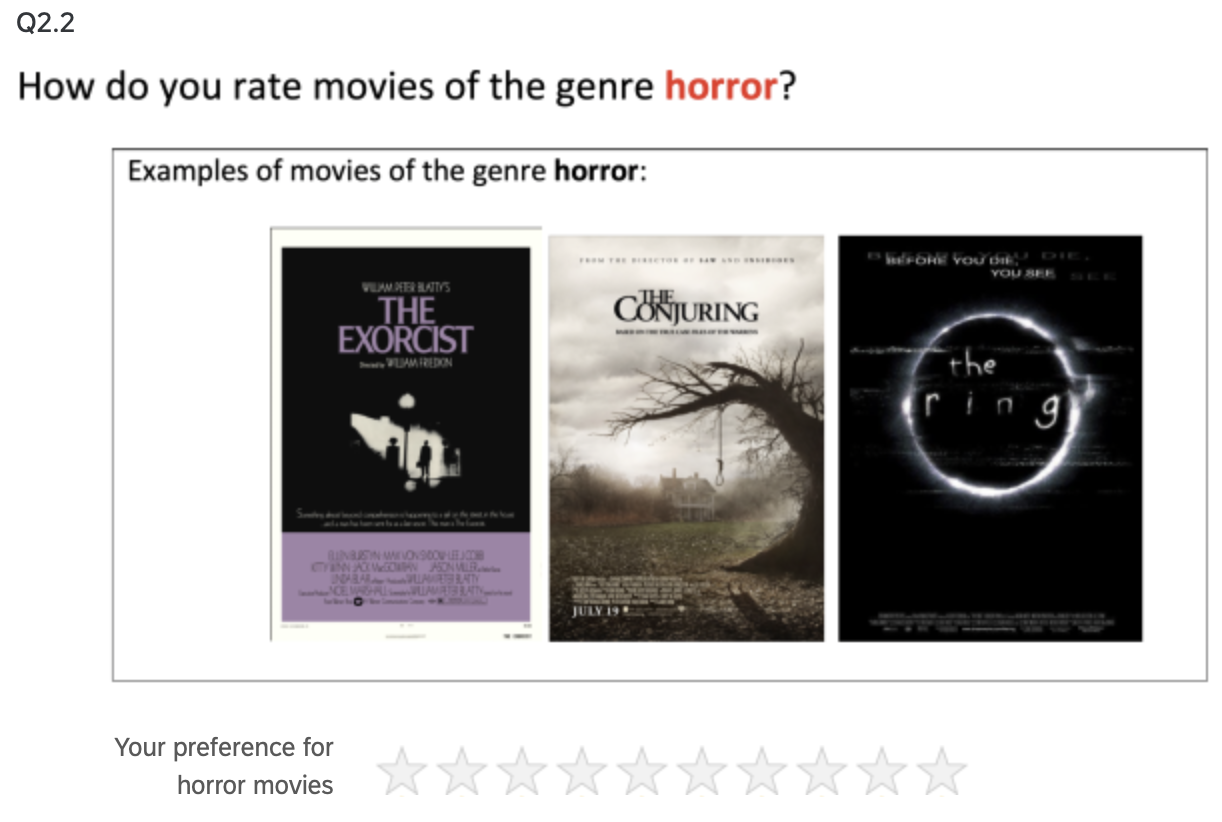}

\includegraphics[width=0.9\textwidth]{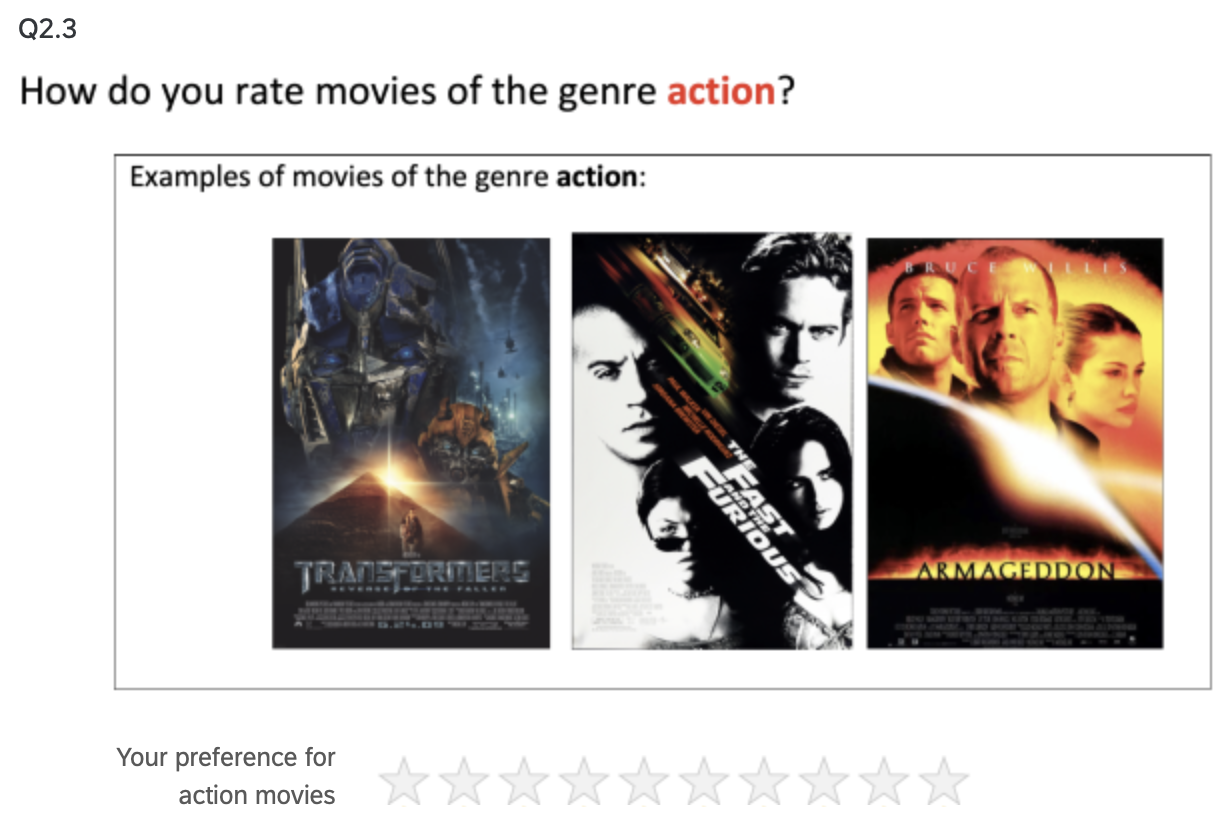}

\includegraphics[width=0.9\textwidth]{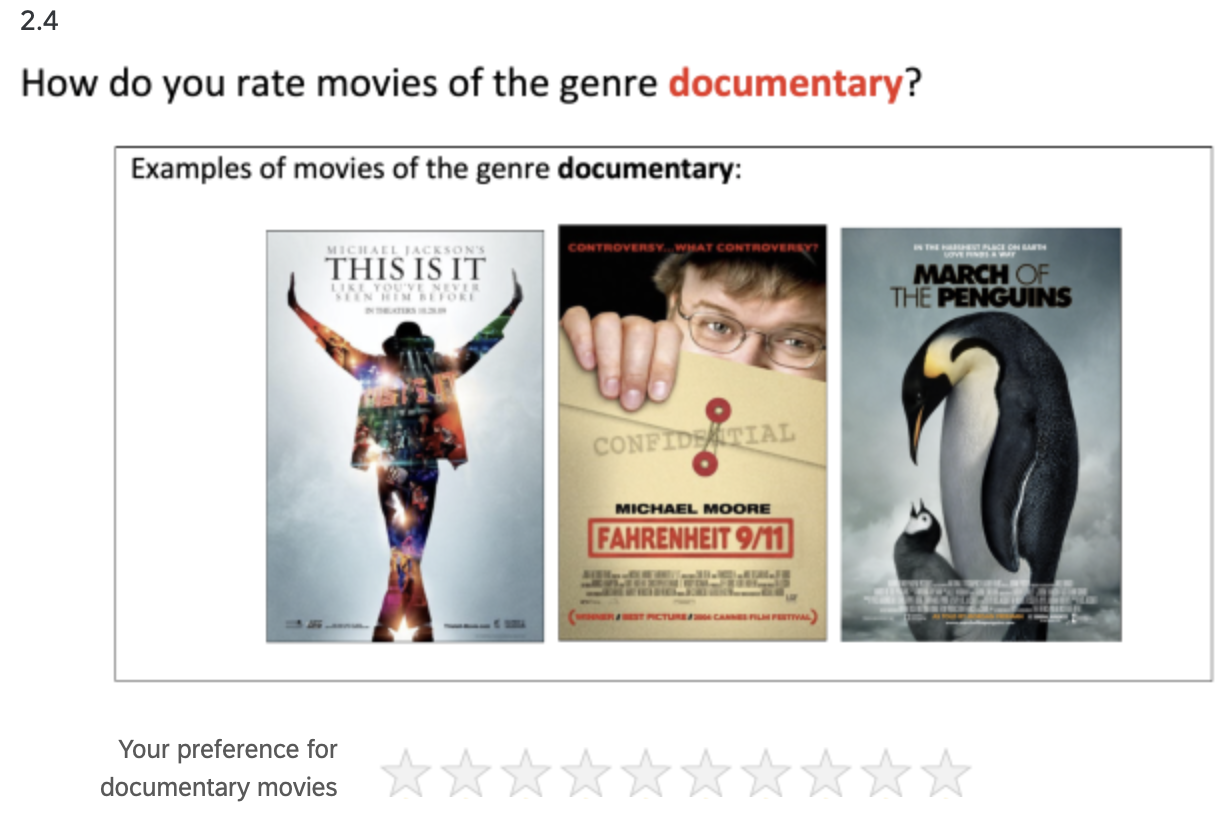}

\includegraphics[width=\textwidth]{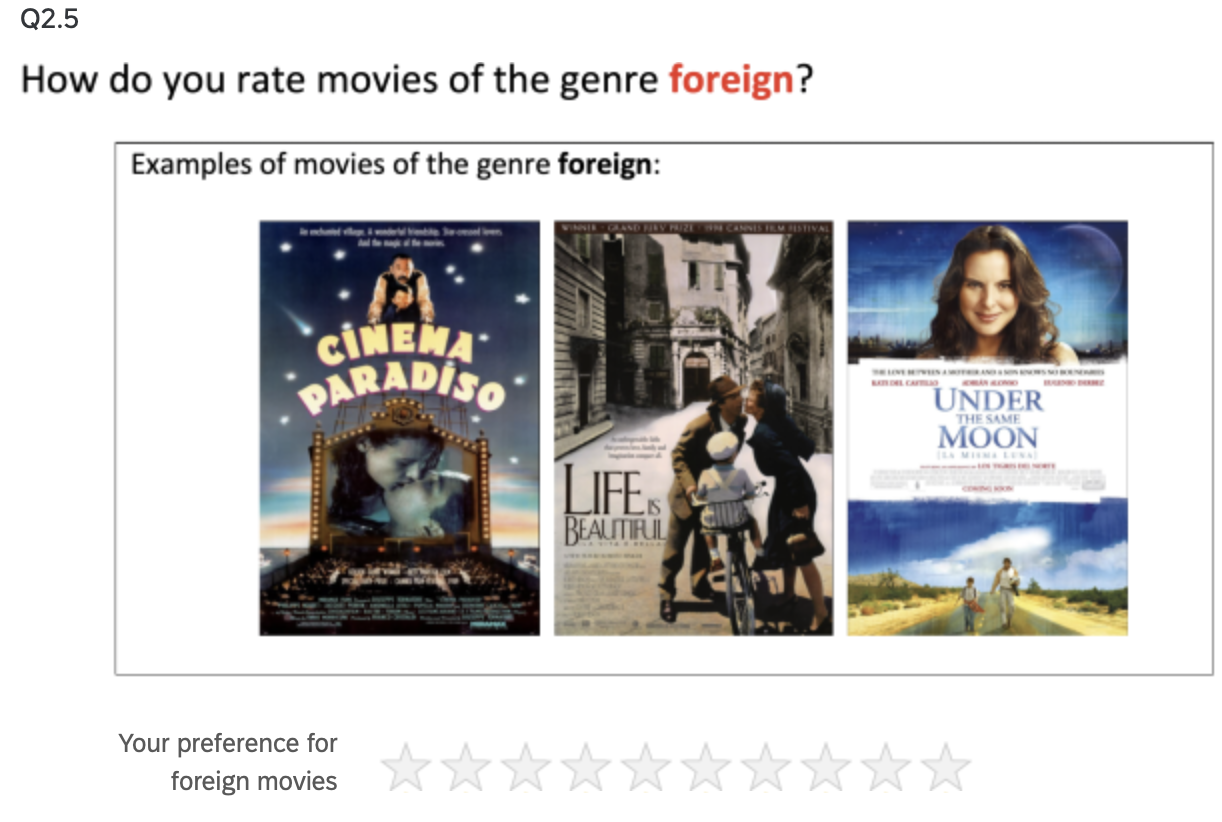}

\includegraphics[width=0.9\textwidth]{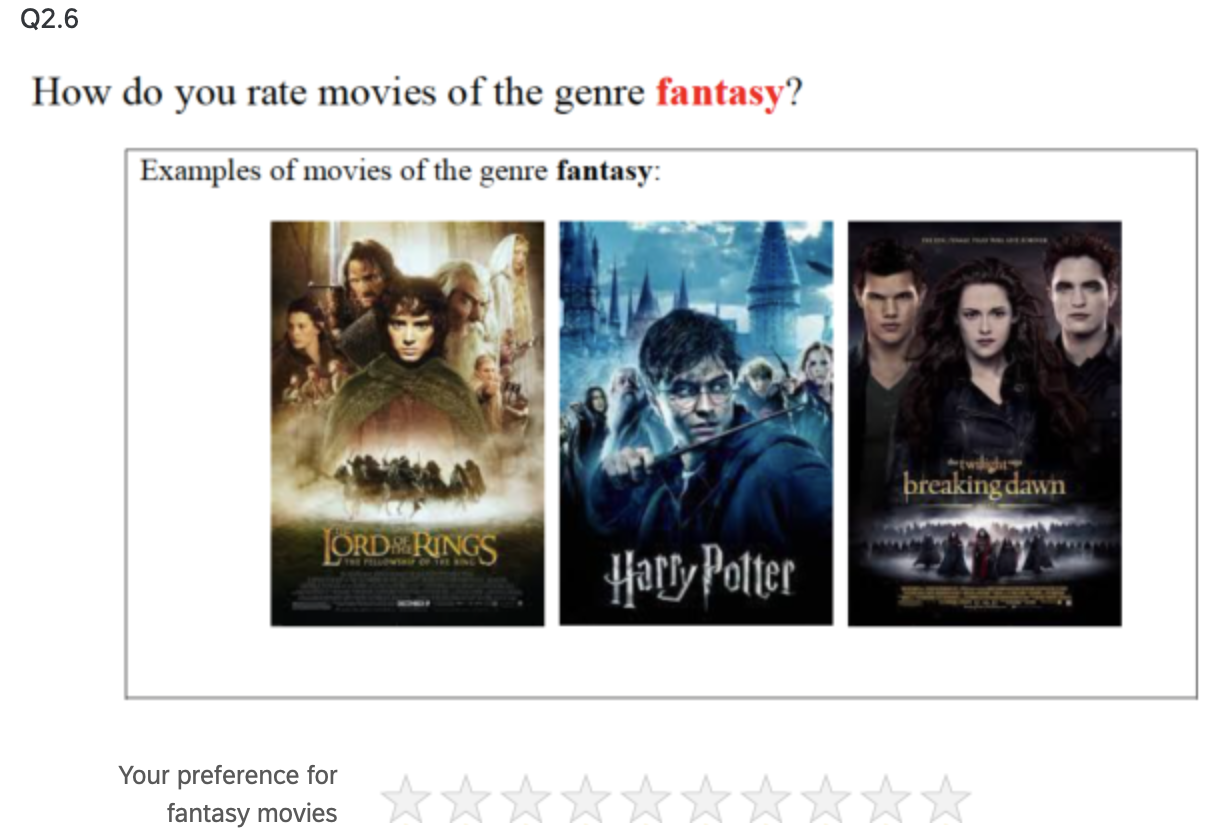}

\includegraphics[width=0.9\textwidth]{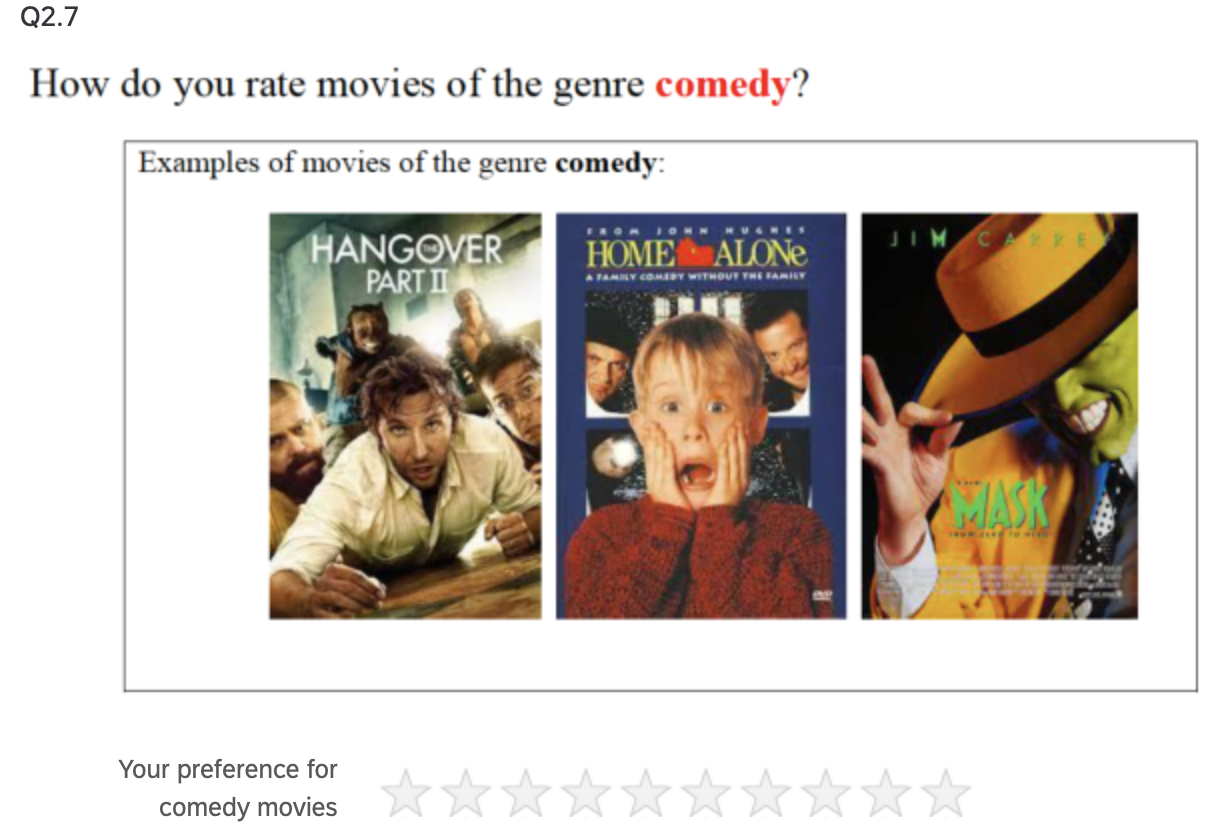}

\includegraphics[width=0.9\textwidth]{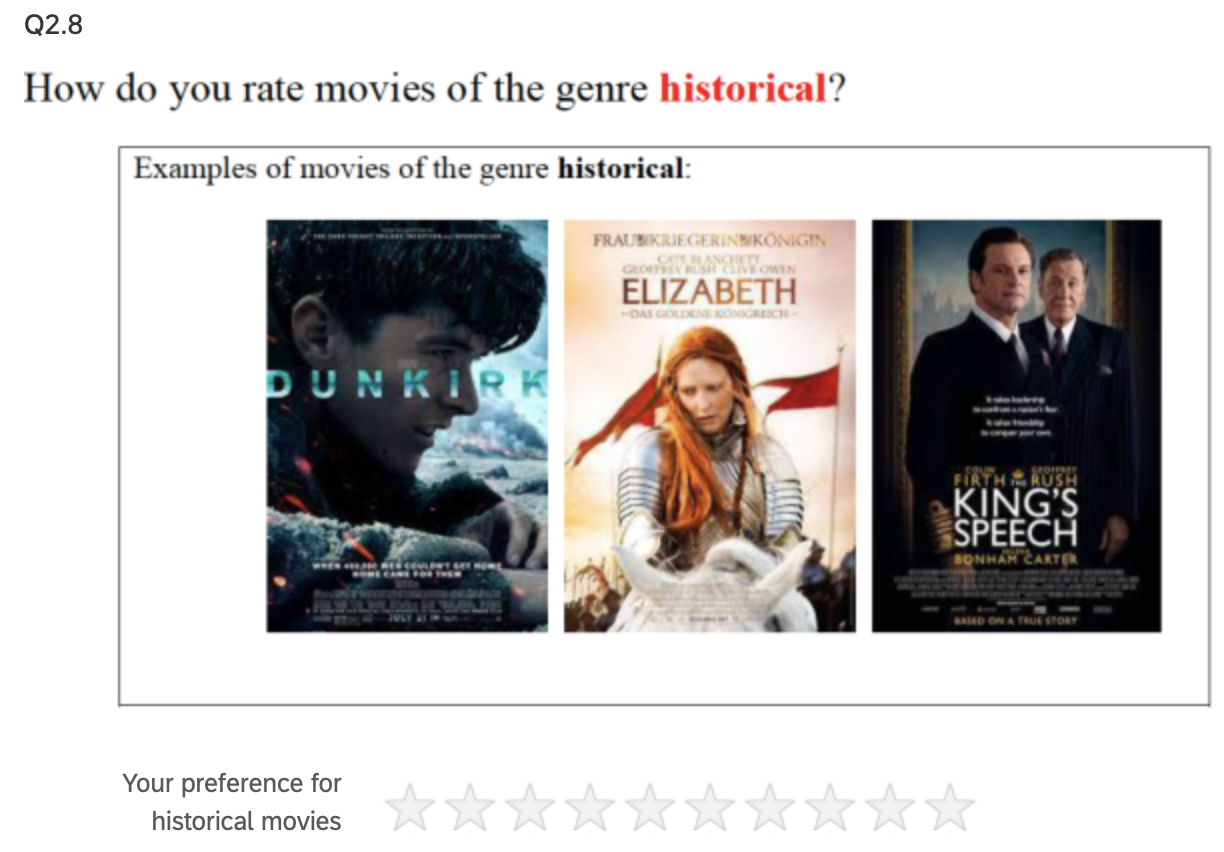}

\includegraphics[width=0.9\textwidth]{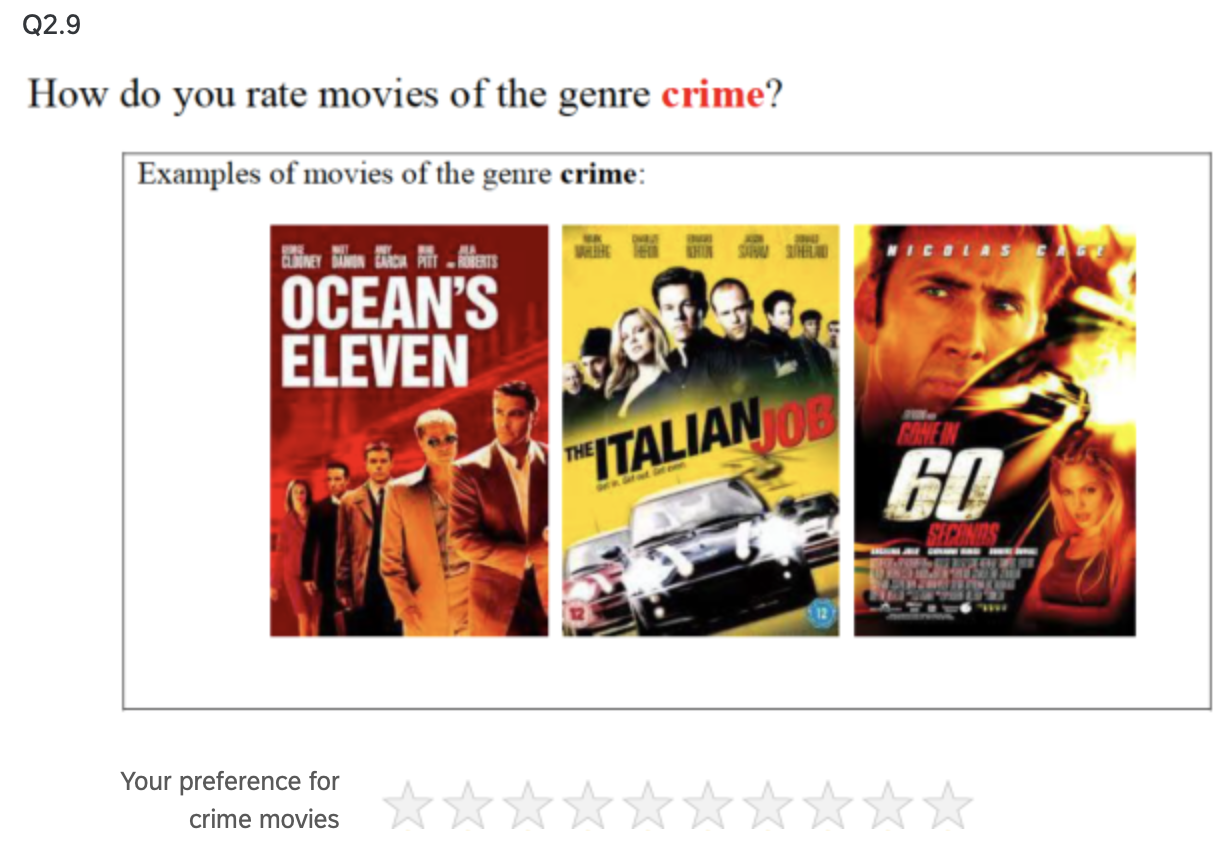}

\includegraphics[width=0.9\textwidth]{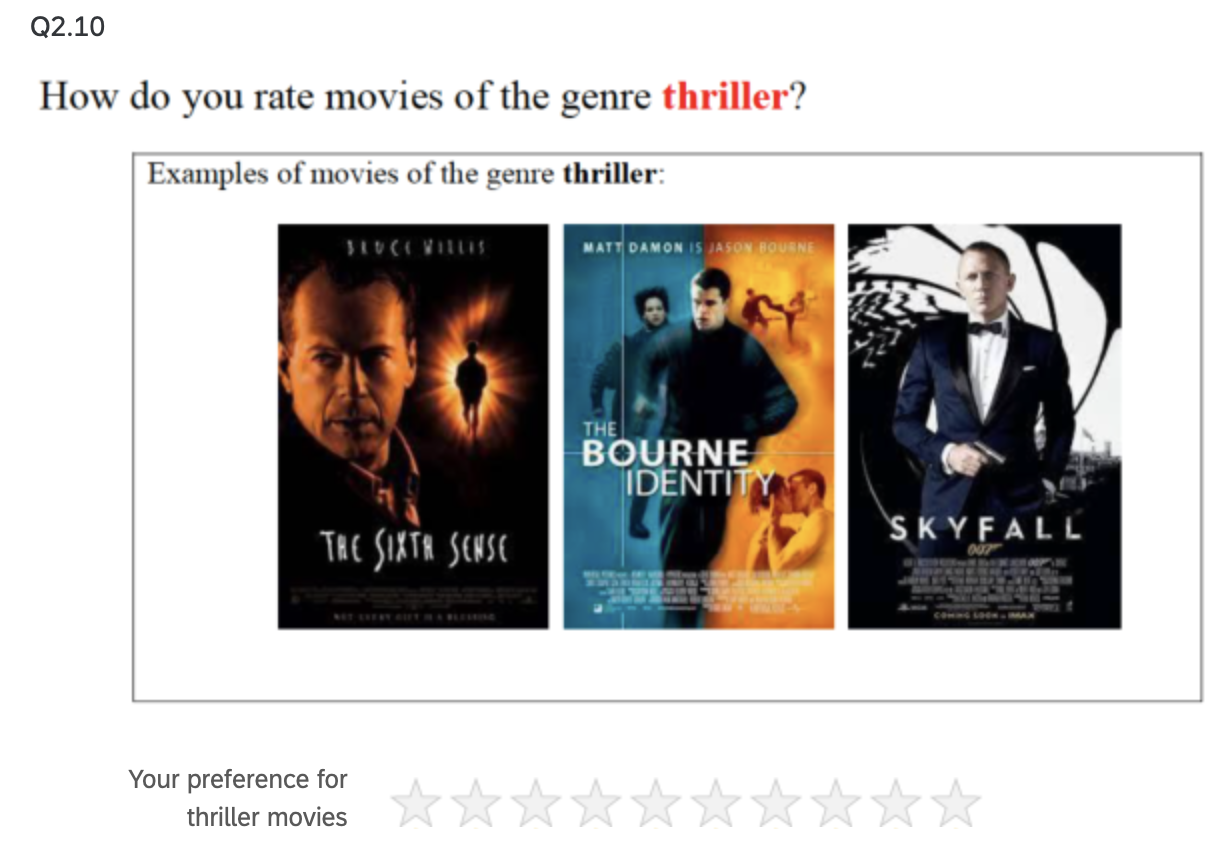}

\vspace{0.6cm}
\textbf{Screen 25}

Imagine a lottery with a 50\% chance of winning £5, and 50\% of winning nothing. Next, you will need to choose whether you would buy this lottery for a corresponding price in each row. 

20 participants who fill out this survey will be chosen randomly. For those, one of the rows below will be chosen randomly. If the participant selected ``do not buy" in that row, he/she will receive a bonus in the form of the corresponding price in the row. If the participant selected ``buy the lottery", she will receive £5 or £0 with a 50\% probability each.

\begin{table}[!htp]\centering
\footnotesize
\begin{tabular}{lrr}\toprule
&Buy the lottery &Do not buy the lottery \\\midrule
Price of £4  	&o	&	o	 \\
Price of £3.8 	 	&o	&	o	 \\	
Price of £3.6    	&o	&	o	 \\
Price of £3.4  		&o	&	o	 \\	
Price of £3.2  		&o	&	o	 \\	
Price of £3  	&o	&	o	 \\	
Price of £2.8 &o	&	o	 \\		
Price of £2.6 &o	&	o	 \\		
Price of £2.4 &o	&	o	 \\		
Price of £2.2 &o	&	o	 \\		
Price of £2 &o	&	o	 \\		
Price of £1.8 &o	&	o	 \\	
Price of £1.6 &o	&	o	 \\	
Price of £1.4 &o	&	o	 \\	
Price of £1.2 &o	&	o	 \\		
Price of £1 &o	&	o	 \\		
Price of £0.8 &o	&	o	 \\	
Price of £0.6 &o	&	o	 \\	
\bottomrule
\end{tabular}
\end{table}

\vspace{1.6cm}

\subsection{Treatments Risk  and Movies}

\textbf{Screen 3}

In the next block, you will answer 10 questions about yourself. 
 
 After these 10 questions, we will offer you an option to buy a lottery from us, which gives you a 50\% chance of winning £5.
 
 After answering the next ten questions, you will have a chance to buy the lottery ticket for a certain price. The price you will face might be personalized by an algorithm based on the statistical relation between other participants' answers to these same questions and how much they were willing to pay for that lottery. The goal of the algorithm is to maximize the revenue obtained from the sale of the lotteries to the participants who choose to buy for the given price.

We will award you with a bonus of £2.20 for answering the questions.

\vspace{0.6cm}

\textbf{Screens 4---14 Risk survey or Movies survey, depending on the treatment}

\vspace{0.6cm}

\textbf{Screen 15}

Your answers to the survey are recorded. Remember that they can influence the price for the lottery on the next screen.
 
 However, for £0.10, you can hide your answers from the algorithm that determines the price. If you hide your answers, you will face an anonymous price, which is set to maximize the revenue from the lottery sales without the information from your survey answers.
 
 Do you want to pay £0.10 and hide your answers (we will deduct it from your bonus of £2.20 for the survey)?
 
 Note that we will inform you of both the anonymous price and the price you would face in case the price was based on your answers at the end of the survey. 
 
o	Pay £0.10 and hide the answers, so the price is not based on my answers  

o	Do NOT hide the answers, so the price can be based on my answers  

\vspace{0.6cm}
\textbf{Screen 16}

Buying: Do you want to buy a lottery with a 50\% chance of winning £5 and a 50\% chance of winning zero? 

 The price is X\footnote{Respective individualized or anonymous price depending on the privacy choice.}

 If you decide to buy, we will deduct the price from the £2.20 bonus you earned for the survey and play out the lottery immediately. If you win, we will add £5 to your bonus.

o	Buy the lottery 

o	Do not buy the lottery 

\vspace{0.6cm}
\textbf{Screen 17}

Belief price range: The personalized prices algorithm, which was generated using answers from other participants and the price that they were willing to pay for the lottery, uses the answers to the survey to determine the price for the lottery. Given that, what do you think are the lowest and highest possible prices that the algorithm generates when considering all possible answers in the survey?

 If your answer is within £0.20 from the correct lowest price, we will add £0.10 to your bonus. 

 If your answer is within £0.20 from the correct highest price, we will add £0.10 to your bonus.

Lowest possible price (slider between 1 and 3.5)

Highest possible price  (slider between 1 and 3.5)

\vspace{0.6cm}
\textbf{Screen 18}

For your information: 

The anonymous price is 1.85

The price based on your answers is Y\footnote{ Individualized price.}

\appendix

\end{document}